%% file: main.tex
\documentclass{article} % For LaTeX2e
\usepackage{iclr2027_conference,times}

\input{math_commands.tex}

\usepackage{hyperref}
\usepackage{url}
\usepackage{booktabs}
\usepackage{multirow}
\usepackage{makecell}
\usepackage{adjustbox}
\usepackage{bbding}
\usepackage{arydshln} % dashed rules: \hdashline, \cdashline{} (load after booktabs/array)
\usepackage[ruled,linesnumbered]{algorithm2e}
\usepackage{dsfont} % \mathds{1} indicator function
\usepackage{tcolorbox}
\tcbuselibrary{skins,fitting,breakable,listings}
\usepackage{xcolor}
\usepackage{graphicx}
\newcolumntype{C}{>{\centering\arraybackslash}X}

\usepackage{listings}
\colorlet{jsonpunct}{red!60!black}
\definecolor{jsondelim}{RGB}{20,105,176}
\colorlet{jsonnumb}{magenta!60!black}
\definecolor{jsonstring}{RGB}{0,128,0}
\lstdefinelanguage{json}{
  basicstyle=\normalfont\ttfamily\fontsize{7}{7.5}\selectfont,
  showstringspaces=false,
  breaklines=true,
  morestring=[b]",
  stringstyle=\color{jsonstring},
  literate=
   *{0}{{{\color{jsonnumb}0}}}{1}
    {1}{{{\color{jsonnumb}1}}}{1}
    {2}{{{\color{jsonnumb}2}}}{1}
    {3}{{{\color{jsonnumb}3}}}{1}
    {4}{{{\color{jsonnumb}4}}}{1}
    {5}{{{\color{jsonnumb}5}}}{1}
    {6}{{{\color{jsonnumb}6}}}{1}
    {7}{{{\color{jsonnumb}7}}}{1}
    {8}{{{\color{jsonnumb}8}}}{1}
    {9}{{{\color{jsonnumb}9}}}{1}
    {:}{{{\color{jsonpunct}{:}}}}{1}
    {,}{{{\color{jsonpunct}{,}}}}{1}
    {\{}{{{\color{jsondelim}{\{}}}}{1}
    {\}}{{{\color{jsondelim}{\}}}}}{1}
    {[}{{{\color{jsondelim}{[}}}}{1}
    {]}{{{\color{jsondelim}{]}}}}{1},
}

\title{MiDashengLM-Spatial: Unifying General Audio Understanding and Spatial Awareness}

\author{Xiaoai Plus, Dasheng Team \\ Xiaomi Inc., China}

\newcommand{\ora}[1]{\textcolor{gray}{#1}}

\iclrfinalcopy % Uncomment for camera-ready version, but NOT for submission.
\begin{document}

\maketitle

\input{tex/00_abstract}
\input{tex/01_introduction}
\input{tex/02_related_works}
\input{tex/03_data_pipeline}
\input{tex/04_method}
\input{tex/05_experiments}
\input{tex/06_conclusion}

\bibliography{iclr2027_conference}
\bibliographystyle{iclr2027_conference}

\newpage
\appendix
\input{appendix/01_dataset_details}
\input{appendix/02_experiment_details}
\input{appendix/03_contributors.tex}
\end{document}

%% file: math_commands.tex
\usepackage{amsmath,amsfonts,bm}

\def\eqref#1{equation~\ref{#1}}
\def\1{\bm{1}}

\DeclareMathAlphabet{\mathsfit}{\encodingdefault}{\sfdefault}{m}{sl}
\SetMathAlphabet{\mathsfit}{bold}{\encodingdefault}{\sfdefault}{bx}{n}

%% file: tex/00_abstract.tex
\begin{abstract}
Large audio-language models (LALMs) have achieved strong performance in general audio understanding, yet most are designed for monaural input and discard the inter-channel cues essential for spatial perception. In contrast, existing spatial audio-language models are purpose-built for spatial tasks and fail to capitalize on the general understanding capabilities of monaural LALMs. We present MiDashengLM-Spatial, the first open-source end-to-end unified audio-language model, to our knowledge, which supports both general audio understanding and spatial awareness within a single architecture. It extends MiDashengLM with a spatial audio encoder, Spatial-Dasheng, integrated through a hierarchical semantic-to-spatial conditioning module that injects intermediate semantic representations into the spatial branch at multiple depths while preserving the original semantic pathway. To provide spatial audio-language supervision at scale, we develop a data synthesis pipeline that renders diverse spatial acoustic scenes with scene-level spatial descriptions and question-answer pairs. Experiments show that Spatial-Dasheng achieves strong performance on sound event localization and detection in real-world scenes, and that MiDashengLM-Spatial substantially outperforms existing LALMs on spatial understanding and reasoning benchmarks. Meanwhile, it remains competitive with state-of-the-art 8B-scale LALMs on diverse monaural benchmarks, demonstrating that spatial awareness can be acquired without compromising general audio understanding. The source code and model checkpoint are available at \href{https://github.com/xiaomi-research/midashenglm-spatial}{\includegraphics[height=1em,keepaspectratio]{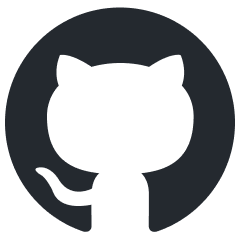}}\footnote{\url{https://github.com/xiaomi-research/midashenglm-spatial}} and \href{https://huggingface.co/mispeech/midashenglm-spatial}{\includegraphics[height=1em,keepaspectratio]{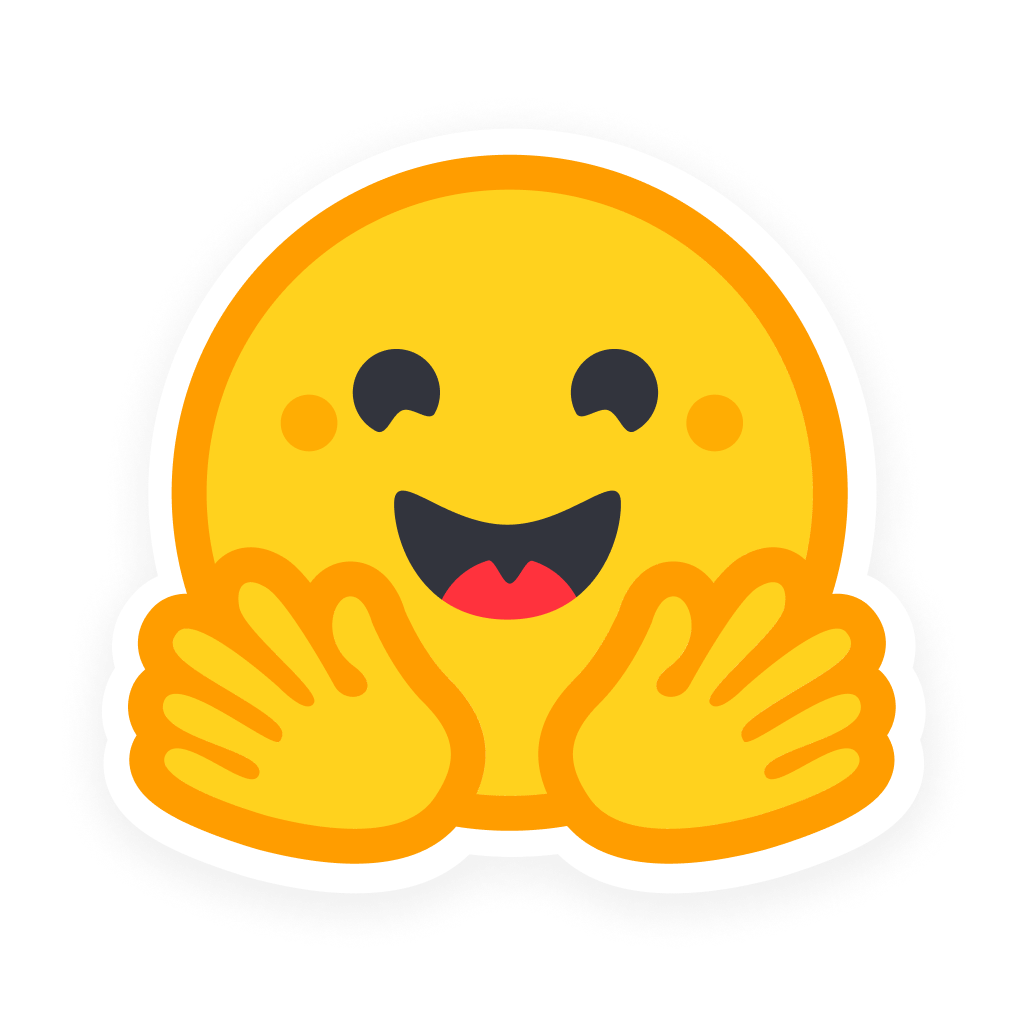}}\footnote{ \url{https://huggingface.co/mispeech/midashenglm-spatial}}.
\end{abstract}

%% file: tex/01_introduction.tex
\section{Introduction}

Humans naturally perceive both what they hear and where sounds originate, with binaural hearing providing essential cues for spatial perception. Inspired by this ability, extensive research has explored sound event localization and detection (SELD) \cite{Adavanne2018_JSTSP,overviewofDCASE,pseldnets,starss23,exp_seldsd}. SELD jointly recognizes sound events and estimates their temporal boundaries and three-dimensional locations, even when multiple sound events overlap. However, conventional SELD systems are typically restricted to predefined event categories and structured outputs, limiting their ability to describe unseen acoustic events, complex spatial relations, and source movements through open-ended natural language.

Although large audio-language models (LALMs) offer flexible natural-language interaction for audio understanding tasks \cite{qwen3omni,gemma4,af-next,dinkel2025midashenglm} and have achieved strong performance across various audio benchmarks, most are developed primarily for monaural audio and do not explicitly model the inter-channel cues required for spatial perception \cite{mmaupro,star-bench}. Recent efforts \cite{BAT,biswas2026owl,dementyev2026phasecoder,SPUR} have extended LALMs with dedicated spatial audio encoders to enable spatial understanding and reasoning. Nonetheless, these methods are designed specifically for spatial tasks, from their model architectures to their data pipelines, and do not fully leverage available monaural LALMs with strong general audio understanding capabilities. Moreover, available monaural audio data is at a substantially larger scale than spatial audio data. Unifying general monaural audio understanding and spatial awareness in a single model remains an open problem.

To address these limitations, we present MiDashengLM-Spatial, which supports both general audio understanding and spatial awareness within a single architecture. MiDashengLM-Spatial extends MiDashengLM \cite{dinkel2025midashenglm} with Spatial-Dasheng through a dual-branch audio encoder equipped with a hierarchical semantic-to-spatial conditioning (HSSC) module. Spatial-Dasheng is a spatial variant of Dasheng \cite{dasheng} designed to capture spatial cues for estimating the direction of arrival (DOA), distance, and motion of overlapping sound sources. HSSC hierarchically conveys intermediate-layer semantic embeddings from the semantic branch to the corresponding layers of the spatial branch, enabling spatial modeling to leverage semantic context at multiple network layers. This design jointly models semantic and spatial information while maintaining a clear functional separation between the two branches.

Given the scarcity of large-scale available spatial audio datasets with rich language annotations, we develop a scalable data synthesis pipeline to construct diverse spatial acoustic scenes. The pipeline generates multi-channel audio clips encompassing environmental sounds, speech, and music, together with scene-level spatial textual descriptions and question-answer (QA) pairs, thereby providing the supervision to align spatial audio representations with language and supporting training for spatial understanding and reasoning.

We evaluate Spatial-Dasheng on publicly available real-scene SELD datasets \cite{starss23,guo2026mrsaudio} and assess MiDashengLM-Spatial on established audio benchmarks that contain dedicated spatial understanding and reasoning subsets \cite{mmaupro,star-bench}. Experimental results show that Spatial-Dasheng achieves superior performance on real-world acoustic scenes, while MiDashengLM-Spatial substantially outperforms existing models on spatial audio understanding without sacrificing its general audio understanding capabilities.

Our contributions are summarized as follows:

\begin{enumerate}
    \item We introduce Spatial-Dasheng, a spatial audio encoder that performs frame-wise detection and localization of overlapping sound events.
    \item We present MiDashengLM-Spatial, the first open-source end-to-end unified model, to our knowledge, which supports both general audio understanding and spatial awareness. Its hierarchical semantic-to-spatial conditioning mechanism integrates MiDashengLM with Spatial-Dasheng while preserving their functional separation.
    \item We develop a scalable data synthesis pipeline that constructs large-scale multi-channel scenes involving environmental sounds, speech, and music, together with rich spatial scene descriptions and question-answer pairs.
    \item Extensive experiments demonstrate that MiDashengLM-Spatial achieves superior performance on spatial audio understanding and reasoning benchmarks while maintaining competitive general audio understanding capabilities, and that Spatial-Dasheng exhibits robust spatial perception and sim-to-real generalization.
\end{enumerate}

%% file: tex/02_related_works.tex
\section{Related Work}
\subsection{Sound Event Localization and Detection}

SELD \cite{Adavanne2018_JSTSP} combines sound event detection (SED) and sound source localization (SSL), and has been a core task in the DCASE\footnote[1]{\url{https://dcase.community}} Challenge since 2019. Existing works can be broadly categorized into single-branch and multi-branch architectures, exemplified by Activity-coupled Cartesian DOA (ACCDOA) \cite{shimada2021accdoa,multiaccdoa} and Event-Independent Network V2 (EINV2) \cite{cao2021,hu2022track,pseldnets} methods, respectively. ACCDOA unifies SED and SSL within a single output representation by encoding event activity into Cartesian DOA vectors, whereas EINV2 adopts separate but interconnected branches that learn partially specialized representations for the two subtasks. These SELD methods demonstrate promising performance in real spatial environments, but they do not provide an open-ended language interface for describing arbitrary spatial acoustic scenes. Our work retains the need for structured spatial audio modeling while extending the output space to language-based spatial reasoning.

\subsection{Large Audio-Language Models and Spatial Extensions}

LALMs \cite{dinkel2025midashenglm,gemma4,moss-audio} connect audio encoders with LLMs to formulate diverse audio tasks as text generation. Through audio-text alignment, these models provide a natural-language interface to the audio modality, supporting general audio understanding and reasoning. Most existing LALMs, however, such as the Qwen-Audio/Omni family \cite{qwen_audio,qwen2_audio,qwen25omni,qwen3omni} and the Audio-Flamingo family \cite{AF3,af-next}, are generally trained on monaural data. When processing multi-channel audio, they typically collapse the channel dimension, for example, by averaging across channels. This operation discards inter-channel cues that are essential for spatial perception.

Prior work on spatial audio-language modeling has explored two main paradigms: encoder-only ALMs and encoder-decoder ALMs. Encoder-only ALMs, e.g., ELSA \cite{ELSA}, Spatial-CLAP \cite{spatial-clap}, SALM \cite{salm}, and CoSTALA \cite{ren2026costala}, learn a joint embedding space for spatial audio and text, enabling spatial attribute modeling and providing spatially grounded representations for downstream tasks such as text-based retrieval, captioning, and editing. In contrast, encoder-decoder ALMs, including BAT \cite{BAT}, OWL \cite{biswas2026owl}, SPUR \cite{SPUR} and PhaseCoder \cite{dementyev2026phasecoder}, augment LALMs with dedicated spatial modeling components, enabling spatial perception while retaining natural-language interaction and producing free-form descriptions or answers. These methods primarily focus on static or clip-level spatial attributes. More recent work \cite{spatia-omni,hyun2026spatio} has begun to model moving sound sources via frame-level source trajectories.

Despite this progress, existing approaches fail to preserve the general audio understanding capabilities acquired through large-scale monaural pre-training: although the aforementioned models typically build upon pre-trained audio encoders, they introduce substantial architectural modifications and extensively fine-tune the encoder on spatially augmented data, compromising its original general understanding ability. Moreover, they typically characterize spatial sound sources with predefined, closed-set labels, limiting their ability to describe complex acoustic scenes in open-ended natural language. Our work fills these gaps by retaining the original audio encoder of the base LALM with minimal modification and synthesizing scene-level spatial audio-language data at scale.

%% file: tex/03_data_pipeline.tex
\section{Data Pipeline}

Although abundant audio resources have been collected and annotated \cite{gemmeke2017audio,librispeech,laion300m,he2024emilia}, most are monaural. In contrast, existing spatial audio datasets are typically recorded with a particular microphone array and thus constrained to a fixed channel configuration \cite{starss23,guo2026mrsaudio,yang2024realman}; multi-channel data with specific array configurations and annotations therefore remains scarce and is largely obtained through simulation \cite{pseldnets,BAT,biswas2026owl,dementyev2026phasecoder}. We construct spatial acoustic scenes at scale, together with corresponding scene-level descriptions and question-answer pairs, for model training.

\subsection{Spatial Acoustic Scene Generation}
\label{sec: spatial_audio_gen}
\begin{figure}[h]
    \centering
    \includegraphics[width=\linewidth]{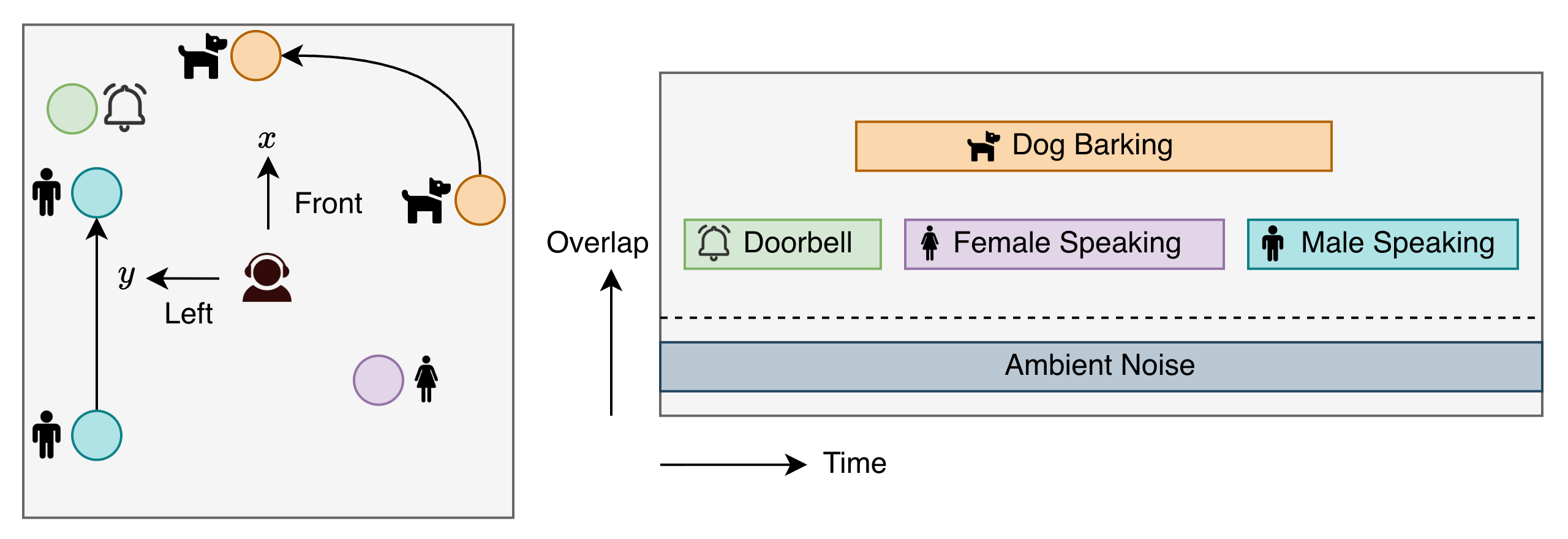}
    \caption{An example of a typical spatial acoustic scene. Left: a top-down view of a room, illustrating the categories, positions, and movement trajectories of sound events. Right: the timeline of the corresponding sound events.}
    \label{fig: spatial_scenes}
\end{figure}

We synthesize spatial audio in two stages. First, we pre-compute a large-scale database of room impulse responses (RIRs) for thousands of randomly sampled rooms using the image source method \cite{ism}, generating RIRs along diverse linear and circular trajectories. Second, dry clips drawn from audio-text corpora, including audio/music captioning and speech transcription datasets, are arranged on a timeline with randomized onsets; each event is then rendered either as a static source at a fixed position by convolving it with the corresponding RIR, or as a moving source by applying time-varying convolution with a sequence of RIRs along its trajectory. Because the scenes are simulated parametrically, complete and exact scene metadata, including source and receiver positions, event timestamps, room dimensions, and reverberation time, is available, enabling frame-level annotations of the azimuth, elevation, and distance of every active source. Figure \ref{fig: spatial_scenes} presents a typical synthetic spatial acoustic scene.

\subsection{Scene-level Spatial Description and Question-Answer Pair Generation}
\label{sec: spatial_caption_gen}
Spatial acoustic scenes require corresponding scene-level descriptions to align spatial audio with language and support spatial understanding, following the training data of general LALMs \cite{dinkel2025midashenglm,af-next}. The spatial metadata described in Sec. \ref{sec: spatial_audio_gen} enables a holistic description of each spatial acoustic scene, specifying what occurs at each moment, where each sound source is located, and how the sources are temporally related.

We first convert the raw numerical spatial information into linguistic descriptors. Static source positions are quantized into directional expressions (e.g., \emph{above} and \emph{front-right}), while moving sources are described by their direction of motion and their start and end positions. The complete mapping from spatial attributes to natural-language descriptors is provided in Appendix \ref{appendix: mapping}. Subsequently, we prompt an LLM (Gemma-4-31B \cite{gemma4} in this work) with the ground-truth spatial metadata and the original audio captions or speech transcriptions for all events in the scene. The LLM synthesizes these event-level descriptions into a coherent narrative of the entire scene. Further details are provided in Appendix \ref{appendix: spatial_caption}. Depending on the source corpus, we adopt two formulations for scene-level description generation: spatial audio captioning for general audio scenes, in which the output describes the semantics of each event along with its spatial and temporal attributes; and spatial speech transcription for speech scenes, in which the output preserves the spoken content while referring to each speaker using spatially grounded expressions. An example description of the scene shown in Figure \ref{fig: spatial_scenes} is provided below:

\begin{tcolorbox}[colback=black!5!white, colframe=black!55!white,
                  fonttitle=\bfseries, fontupper=\footnotesize]
The scene begins with a doorbell ringing in front-left of the listener, overlapping with a dog barking as it moves counter-clockwise from the listener's front-right and below to directly in front and below. Next, a woman statically positioned behind and to the listener's right asks, ``Who is at the door?" Finally, a man moves from the back-left to the front-left of the listener and says, ``I'll take a look."
\end{tcolorbox}

Rather than prompting an LLM to generate questions and validating them post hoc, we deterministically construct multiple-choice spatial QA pairs from the spatial metadata, ensuring that every answer can be directly derived from the underlying scene facts. Specifically, an oracle module first extracts event-level spatial facts from the coordinates, timestamps, and semantic descriptions of each source. These facts are then used to instantiate a set of manually designed question templates, yielding QA pairs whose correct answers and distractors are both verified against the extracted facts. The resulting QA pairs cover three categories: source localization (localizing a given source or identifying the source in a given direction), motion trajectory perception, and group-level spatio-temporal relations. A fourth category, spatially grounded content understanding, requires joint reasoning over event semantics and spatial context, which lies beyond what fixed question templates can express. Therefore, we synthesize it with a different strategy. We render these scenes from dialogue corpora with speaker-identity annotations (e.g., AliMeeting \cite{yu2022alimeeting}), as multi-turn dialogues exhibit coherent semantic logic and rich inter-speaker dependencies. An LLM then generates QA pairs from the dialogue content, with each speaker mention deterministically remapped to a spatial-direction referent derived from the metadata. Finally, an LLM rephrases all QA pairs to increase linguistic diversity while preserving their verified semantics. More details on the spatial QA pairs are provided in Appendix~\ref{appendix: spatial_qa}.

%% file: tex/04_method.tex
\section{Method}

\begin{figure}
    \centering
    \includegraphics[width=\linewidth]{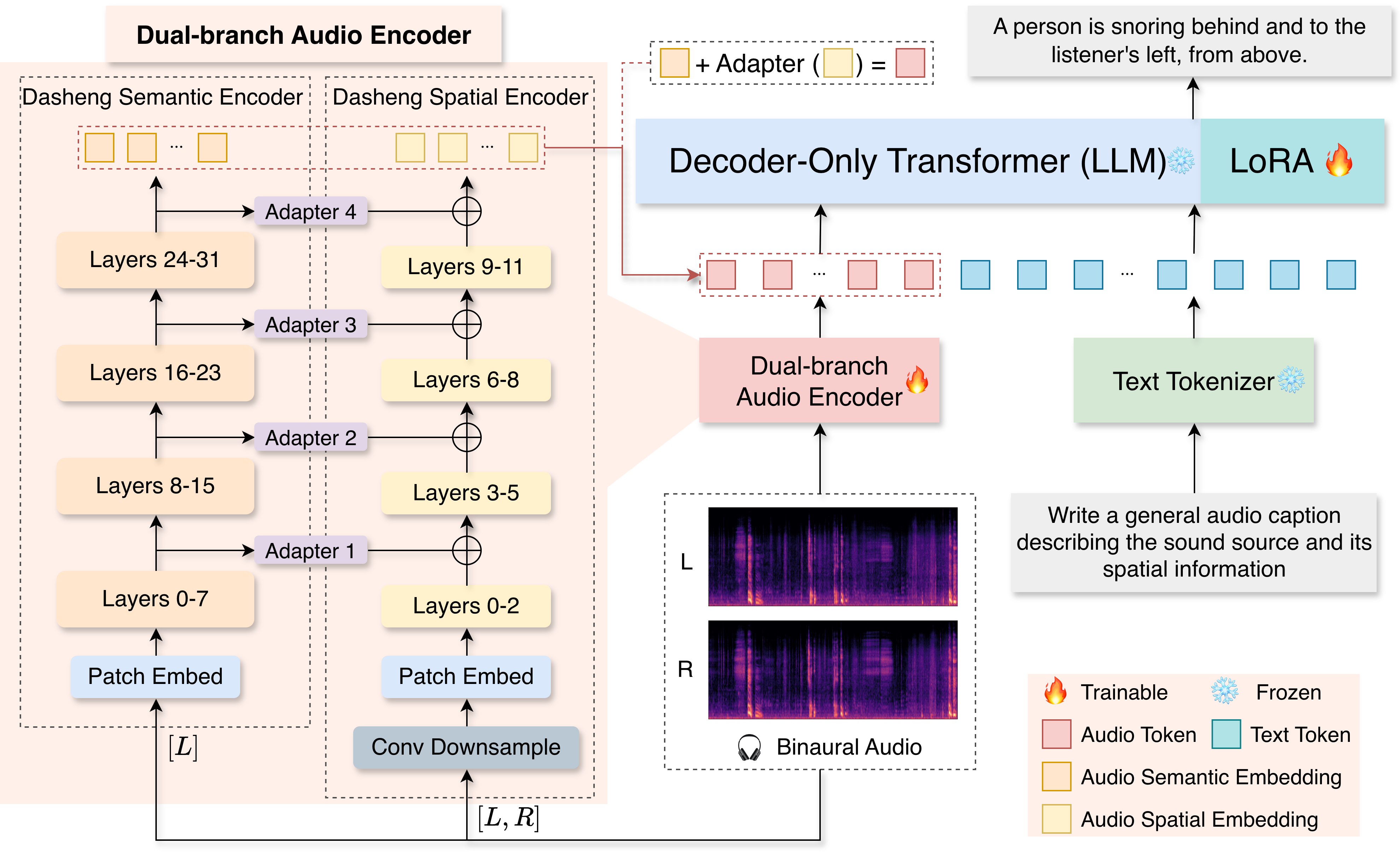}
    \caption{The overall architecture of MiDashengLM-Spatial. The left part shows the details of the dual-branch audio encoder.}
    \label{fig: midashenglm_spatial}
\end{figure}

Figure \ref{fig: midashenglm_spatial} presents the overall architecture of MiDashengLM-Spatial. Built upon MiDashengLM \cite{dinkel2025midashenglm}, our model incorporates Spatial-Dasheng through a hierarchical semantic-to-spatial conditioning (HSSC) module. The 32-layer Dasheng Semantic Encoder, corresponding to MiDashengLM's original audio encoder, processes monaural audio to extract semantic representations, while the 12-layer Dasheng Spatial Encoder, instantiated as Spatial-Dasheng, processes multi-channel audio to capture spatial cues.

\subsection{Spatial-Dasheng}

The architecture of Spatial-Dasheng is illustrated by the Dasheng Spatial Encoder in Figure \ref{fig: midashenglm_spatial}. Spatial-Dasheng is built upon Dasheng-Base \cite{dasheng}, a frame-level Vision Transformer (ViT) \cite{vit} pre-trained on monaural audio using the Masked Autoencoder (MAE) objective \cite{mae}. To enable spatial audio processing, we introduce a lightweight CNN-based fusion module that aggregates multi-channel input features into single-channel features. The resulting downsampled features are compatible with the standard input to Dasheng-Base.

To enable the localization and motion tracking of overlapping sources, we adopt a SELD-style pre-training paradigm with the multi-ACCDOA output format \cite{multiaccdoa,exp_seldsd}. The multi-ACCDOA output format comprises $N$ tracks, with each detecting at most a single event with its corresponding location. The learning target for the $n$-th track at time frame $t$ is formulated as $\mathbf{O}_{nt}=[\mathbf{P}_{nt}, d_{nt}]$, where $\mathbf{P}_{nt} \in \mathbb{R}^3$ and $d_{nt} \in (0, +\infty)$ denote the Activity-coupled Cartesian Direction of Arrival (ACCDOA) vector and source distance, respectively. Specifically, $\mathbf{P}_{nt}=a_{nt}\mathbf{R}_{nt}$ jointly encodes the source activity $a_{nt} \in \{0, 1\}$ and the DOA vector $\mathbf{R}_{nt}=(x_{nt},y_{nt},z_{nt})$ on the unit sphere. When $a_{nt}=1$, the corresponding sound event is active and $\lVert \mathbf{R}_{nt} \rVert = 1$.

However, track-wise approaches may suffer from misalignment between the target and the prediction. To address this issue, we adopt permutation-invariant training (PIT) \cite{pit} during training. The PIT loss for the track-wise output format is defined as
\begin{equation}
\mathcal{L}_{\mathrm{PIT}}=\sum_{t}\min_{\alpha \in \mathrm{Perm}(t)}
\left\{
\lambda\cdot\ell_{\mathrm{ACCDOA}}(t,\alpha)
+
(1-\lambda)\cdot \ell_{\mathrm{Dist}}(t,\alpha)
\right\},
\end{equation}
where $\alpha\in \mathrm{Perm}(t)$ denotes a possible permutation that yields the minimum loss, and $\lambda$ is a weighting factor that balances the ACCDOA loss and the distance loss. In this work, we employ the mean squared error (MSE) as $\ell_{\mathrm{ACCDOA}}$ and the mean absolute percentage error (MAPE) as $\ell_{\mathrm{Dist}}$.

\subsection{HSSC: Hierarchical Semantic-to-Spatial Conditioning}

Spatial-Dasheng provides spatial perception by detecting sound-event activities and estimating the positions and motion trajectories of overlapping sources. In overlapping-source scenarios, extracting spatial cues alone is insufficient, as these cues must be associated with their corresponding semantic cues. To establish this semantic-spatial association and bridge source-level spatial perception with language-based spatial understanding, we introduce an HSSC module that integrates Spatial-Dasheng into MiDashengLM, as depicted in the left part of Figure \ref{fig: midashenglm_spatial}. Specifically, HSSC uses intermediate representations from the Dasheng Semantic Encoder to condition spatial feature extraction at multiple depths of the Dasheng Spatial Encoder:
\begin{equation}
    E_{3k-1}^{\mathrm{Spa}}
    \leftarrow
    E_{3k-1}^{\mathrm{Spa}}
    +
    \mathrm{Adapter}_{k}\left(E_{8k-1}^{\mathrm{Sem}}\right),
    \quad k \in \{1,2,3,4\}.
\end{equation}
Here, $E_{8k-1}^{\mathrm{Sem}}$ denotes the intermediate semantic representation extracted from the semantic branch. It is projected through a dedicated merge adapter and then injected into the corresponding spatial representation $E_{3k-1}^{\mathrm{Spa}}$. By conditioning spatial representations on sound-event semantics throughout the audio encoder hierarchy, HSSC facilitates the association of spatial cues with their corresponding semantic cues, particularly when multiple sources overlap. The conditioned spatial representations are subsequently passed to the following layers of the spatial encoder for further processing.

Each merge adapter primarily consists of an MLP projection. Since the dual-branch audio encoders have asymmetric architectures, semantic-to-spatial conditioning is performed every eight layers in the semantic branch and every three layers in the spatial branch, resulting in four hierarchical conditioning modules. This asymmetric design establishes a unidirectional conditioning pathway during the forward pass, whereby intermediate representations from the pretrained semantic branch are injected into the spatial branch, while the newly introduced and task-specific spatial representations are not injected back into the semantic feature hierarchy. In this way, the spatial branch can leverage semantic information for spatial analysis while the original forward computation pathway of the semantic branch remains structurally unchanged.

After the final layers of the two branches, the semantic and spatial embeddings, i.e., $E^{\mathrm{Sem}}$ and $E^{\mathrm{Spa}}$, respectively, are aligned and fused into unified audio tokens $E_{\mathrm{Audio}}$ through an additional adapter:
\begin{equation}
    E_{\mathrm{Audio}}
    =
    E^{\mathrm{Sem}}
    +
    \mathrm{Adapter}\!\left(E^{\mathrm{Spa}}\right),
\end{equation}
The resulting audio tokens are subsequently provided to the LLM for audio understanding.

\subsection{Training Curriculum}

Training directly on complex spatial-understanding tasks requires the model to acquire spatial perception, grounding, and understanding capabilities. Accordingly, we train MiDashengLM-Spatial using a three-stage curriculum designed to develop these abilities progressively.

In Stage I, Spatial-Dasheng is pre-trained on the synthetic SELD datasets described in Sec. \ref{sec: spatial_audio_gen}. The model is trained to predict frame-level event activities, directions of arrival (DOAs), and source distances, without modeling sound categories, thereby acquiring basic spatial perception capabilities. In Stage II, the Dasheng Spatial Encoder of MiDashengLM-Spatial is initialized from the pre-trained Spatial-Dasheng model and fine-tuned on the spatial audio-text paired datasets described in Sec. \ref{sec: spatial_caption_gen}. These datasets include both event-level and scene-level data, facilitating the alignment of spatial attributes with natural language. In Stage III, MiDashengLM-Spatial is fine-tuned on the complete training set, which includes both monaural and binaural corpora, with an emphasis on spatial QA pairs. This final stage equips MiDashengLM-Spatial with comprehensive spatial-understanding capabilities while preserving a balance between general audio understanding and spatial awareness.

%% file: tex/05_experiments.tex
\section{Experiments}

\subsection{Implementation Details}

Our spatialized audio corpora are collected from two sources: audio captioning and speech transcription datasets. The former includes captioned audio, speech, and music corpora, such as ACAVCaps \cite{acavcaps}, LP-MusicCaps \cite{LP-MUSICCAPS}, and AudioCaps \cite{audiocaps}, whereas the latter focuses only on transcribed speech, as in LibriSpeech \cite{librispeech}. To construct the spatial acoustic scene dataset, following the criteria outlined in \cite{pseldnets}, we first identify samples containing only a single sound event using Qwen3-32B \cite{qwen3} and Qwen3-Omni-30B-A3B-Instruct \cite{qwen3omni}, based on the corresponding caption and the audio clip. Subsequently, we spatialize each sound event by convolving it with a simulated room impulse response (RIR) generated using Pyroomacoustics \cite{pyroomacoustics} and a head-related transfer function (HRTF) selected from the ARI database\footnote[2]{\url{https://www.oeaw.ac.at/isf/outreach/software/hrtf-database}}, thereby synthesizing binaural audio. The monaural audio corpora used in Stage III are consistent with those employed in \cite{dinkel2025midashenglm}.

In total, we synthesize 1 million audio clips and 6 million question-answer (QA) pairs, corresponding to approximately 13,000 hours of binaural audio. Each clip has a duration of 30-60 seconds and is accompanied by one generated scene-level spatial description, a set of event-level spatial descriptions corresponding to the number of sound events, and 6 spatial QA pairs. The maximum number of overlapping sources is 3. 

All audio clips are resampled to 16 kHz and truncated to a maximum duration of 60 seconds. For each binaural audio, we extract a two-channel, 64-bin log-mel spectrogram using a 512-point Hann window and a hop length of 160 points. For non-binaural audio (i.e., from monaural corpora), we duplicate the first-channel spectrogram along the channel dimension to match the two-channel input format required by MiDashengLM-Spatial. All model configurations and training strategies otherwise follow those of MiDashengLM \cite{dinkel2025midashenglm}.

\subsection{Spatial-Dasheng Performance on SELD Tasks}

We evaluate Spatial-Dasheng against PSELDNets \cite{pseldnets} and Spatial-AST \cite{BAT} on MRSSound \cite{guo2026mrsaudio}, EasyCom \cite{donley2021easycom}, the stereo version of STARSS23 \cite{starss23,dcase2025task3}, and our synthetic binaural test set. PSELDNets achieves state-of-the-art (SOTA) performance on multiple SELD benchmarks through pre-training, while Spatial-AST serves as the spatial audio encoder for BAT \cite{BAT}. Since stereo recordings provide limited cues for resolving up-down and front-back ambiguities, our evaluation on STARSS23 considers only left-right azimuth estimation.

Following the official SELD metrics \cite{dcase2025task3}, we report the localization-dependent F-score ($\mathrm{F}_{45^\circ/1}$), computed using a $45^\circ$ DOA error threshold and a 100\% relative distance error (RDE) threshold, and the classification-dependent DOA error ($\mathrm{DOAE}_{\mathrm{CD}}$) and RDE ($\mathrm{RDE}_{\mathrm{CD}}$), conditioned on correctly detected events. 
% An effective system should demonstrate a high $\mathrm{F}_{45^\circ/1}$, a low $\mathrm{DOAE}_{\mathrm{CD}}$, and a low $\mathrm{RDE}_{\mathrm{CD}}$.

\input{table/01_seld_results}

Table \ref{tab: res_seld} compares the performance of the evaluated models across the SELD benchmarks. All training datasets and SynthBinaural-Test are synthetically generated, whereas the other three benchmarks comprise recordings collected in real-world environments. Spatial-AST underperforms the other two models on most metrics, likely because its architecture is designed for clip-level prediction and is therefore less suited to fine-grained frame-level localization. In contrast, Spatial-Dasheng substantially outperforms both baselines on EasyCom and MRSSound, while achieving performance competitive with PSELDNets on STARSS23 and SynthBinaural-Test. Notably, despite being trained exclusively on binaural data, Spatial-Dasheng generalizes well to the stereo version of STARSS23. This result suggests that the model can effectively exploit interaural level differences (ILD) to localize sound events. Overall, these results demonstrate the strong spatial perception capability and sim-to-real generalization of Spatial-Dasheng across diverse acoustic conditions.

\subsection{Spatially Grounded Evaluation}
\label{sec: spatially_grounded_eval}

We conduct detailed evaluations on synthetic test sets to assess whether MiDashengLM-Spatial grounds the semantic, spatial, and temporal attributes of the input audio in its textual outputs. We consider two complementary evaluation settings. For spatial audio--text alignment, we evaluate the model on spatial audio captioning and spatial speech transcription test sets (see Appendix \ref{appendix: spatial_align_eval}). We reverse-parse the generated descriptions into structured event-level attributes, match them to reference events based on semantic content, and evaluate event recall, motion-state agreement, static localization, motion-trajectory prediction, and temporal-order consistency. For spatial QA (see Appendix \ref{appendix: spatial_aqa_eval}), we assess the model's spatio-temporal and semantic reasoning across question categories involving source localization, motion-trajectory perception, and group-level spatio-temporal relations. We further analyze how integrating Spatial-Dasheng into MiDashengLM affects localization accuracy (see Appendix \ref{appendix: localization_prf_diff}). Together, these experiments provide a fine-grained assessment of the model's spatial grounding capabilities, complementing the comparisons with existing systems on real-world benchmarks presented in the following section.

Overall, MiDashengLM-Spatial performs well on coarse-grained spatial attributes and relatively simple scenes, but remains less effective at capturing fine-grained spatial details and reasoning over complex scenes with multiple overlapping sources. The spatial audio-text alignment results in Table \ref{tab: spatial_grounded} support this observation: on the non-overlapping (OV1) subsets, the model achieves matched-event recall above $70\%$, motion-state accuracy above $88\%$, and temporal-order consistency above $96\%$, while its exact-match accuracy for motion trajectories remains below $20\%$. A similar pattern emerges in spatial QA. As shown in Table \ref{tab: spatial_qa_results}, the model substantially exceeds the $28.8\%$ random-guess baseline, achieving overall accuracies of $80.6\%$ and $66.9\%$ on OV1 and OV2, respectively, but performs the worst on Spatial Counting, which requires integrating spatial, temporal and semantic information across the entire scene. Finally, the localization comparison in Table \ref{tab: localization_prf} validates the effectiveness of integrating Spatial-Dasheng into MiDashengLM. The integration preserves the model's underlying localization capability and improves the three-axis exact-match F-score from $44.7\%$ to $50.8\%$, suggesting that event-level language modeling may help aggregate frame-level localization cues into more accurate event-level predictions.

\subsection{Evaluation on Spatial and General Audio Understanding Benchmarks}

Table \ref{tab: spatial_benchmarks} reports results on two spatial-audio benchmarks: the spatial-audio subset of MMAU-Pro \cite{mmaupro} and the Spatial Reasoning (SR) subset of STAR-Bench \cite{star-bench}. Both benchmarks consist of audio recorded in real-world environments. Although most of the evaluated multimodal LLMs, including both closed- and open-source models, achieve promising scores on the full MMAU-Pro benchmark, their performance on the two spatial-understanding subsets remains substantially below that of MiDashengLM-Spatial. BAT \cite{BAT}, the only existing open-source binaural spatial LALM, likewise offers no clear advantage on either benchmark. Its narrow-domain training data may limit its instruction-following ability, causing it to struggle with the diverse prompts in these benchmarks despite receiving binaural input. By contrast, the other evaluated models discard multichannel information during audio feature extraction. The performance margin of MiDashengLM-Spatial over the random-guess baselines is notably larger on STAR-Bench than on MMAU-Pro. This difference reflects the distinct emphases of the two benchmarks: STAR-Bench directly targets spatial reasoning, whereas MMAU-Pro places greater demands on the semantic interpretation of acoustic scenes and spoken dialogue. On MMAU-Pro, strong general-purpose models may therefore exploit semantic priors to infer answers without fully grounding them in the spatial information provided by the audio. Consequently, standard benchmark accuracy alone is insufficient to establish whether a model's predictions genuinely rely on binaural spatial cues.

\input{table/02_spatial_benchmarks}

To further test whether model predictions are grounded in binaural cues, we conduct a channel-swap consistency evaluation. Specifically, we select QA items whose correct answers depend solely on whether a sound source is located to the listener's left or right, and evaluate each item using both the original audio and a counterpart with the left and right channels swapped. Channel swapping, which is widely used for post-processing and data augmentation in SELD \cite{nguyen2021salsa,mazzon2019first}, reverses the perceived left-right locations of sound sources in a listener-centered reference. A spatially grounded model should therefore change its answer accordingly. Table \ref{tab: spatial_benchmarks} reports the channel-swap all-correct rate (ACR), defined as the proportion of items answered correctly under both the original and channel-swapped configurations. This metric is stricter than accuracy under a single configuration because it requires both correct reasoning and sensitivity to binaural spatial cues. The channel-swap ACRs of the closed-source models are close to the random-guess baseline, whereas those of the open-source models are nearly zero. This result is expected because most models are designed for monaural audio and average the input waveforms across channels, making their representations invariant to channel order. Deterministic greedy decoding further contributes to the near-zero ACR of the open-source models: a channel-insensitive model produces identical answers for both inputs and therefore cannot be correct under both configurations once channel swapping reverses the target left-right relation. In contrast, MiDashengLM-Spatial substantially outperforms all baselines, indicating that its spatial inferences are grounded in binaural cues rather than language priors, monaural audio cues, or option-position biases.

\input{table/03_general_benchmarks}

To assess whether incorporating binaural spatial modeling affects general audio understanding, we evaluate MiDashengLM-Spatial on monaural-audio benchmarks covering audio question answering, ASR, and audio captioning. We compare it with advanced 8B-scale LALMs, including Qwen2.5-Omni-7B \cite{qwen25omni}, MOSS-Audio-8B-Instruct \cite{moss-audio}, and Audio-Flamingo-Next-Instruct \cite{af-next}, as well as its monaural counterpart, MiDashengLM \cite{dinkel2025midashenglm}. As shown in Table \ref{tab: general_benchmarks}, MiDashengLM-Spatial performs comparably to MiDashengLM and is competitive with state-of-the-art systems of a similar scale across all three task categories. It even outperforms its monaural counterpart on several benchmarks, including multilingual ASR on GigaSpeech2 and music captioning on MusicCaps.

We further analyze two ablation variants reported in the last column of Table \ref{tab: general_benchmarks}: (b) replacing the unidirectional HSSC with bidirectional connections in the audio encoder, i.e., adding an additional spatial-to-semantic injection pathway, and (c) training exclusively on binaural data, without monaural corpora. We observe that, compared with the bidirectional variant (b), our unidirectional HSSC better preserves general audio understanding, outperforming it on 13 of the 18 benchmark settings. Moreover, even when trained solely on spatial data, variant (c) retains most of the original capability of MiDashengLM. This is particularly evident on AISHELL-2: although our spatial training corpora contain no Chinese speech transcription data, variant (c) remains close to the full model. Together, these results indicate that MiDashengLM-Spatial enables spatial perception to emerge while retaining the model's fundamental audio understanding capability.

%% file: table/01_seld_results.tex
\begin{table}[h]
\caption{Comparison of different models on SELD tasks. Higher values of $\mathrm{F}_{45^{\circ}/1}$ and lower values of $\mathrm{DOAE}_{\mathrm{CD}}$ and $\mathrm{RDE}_{\mathrm{CD}}$ indicate better performance.}
% \caption{Comparison of different models on SELD tasks.}
\centering
\label{tab: res_seld}
\begin{adjustbox}{width=\columnwidth,center}
\begin{tabular}{lcccccccccccc}
\toprule
\multirow{2}{*}{Model} & \multicolumn{3}{c}{EasyCom} & \multicolumn{3}{c}{STARSS23} & \multicolumn{3}{c}{MRSSound} & \multicolumn{3}{c}{SynthBinaural-Test}\\
\cmidrule(lr){2-4}\cmidrule(lr){5-7}\cmidrule(lr){8-10}\cmidrule(lr){11-13}
& $\mathrm{F}_{45^{\circ}/1}$ & $\mathrm{DOAE}_{\mathrm{CD}}$ & $\mathrm{RDE}_{\mathrm{CD}}$ & $\mathrm{F}_{45^{\circ}/1}$ & $\mathrm{DOAE}_{\mathrm{CD}}$ & $\mathrm{RDE}_{\mathrm{CD}}$ & $\mathrm{F}_{45^{\circ}/1}$ & $\mathrm{DOAE}_{\mathrm{CD}}$ & $\mathrm{RDE}_{\mathrm{CD}}$ & $\mathrm{F}_{45^{\circ}/1}$ & $\mathrm{DOAE}_{\mathrm{CD}}$ & $\mathrm{RDE}_{\mathrm{CD}}$\\
\midrule
Spatial-AST  & 6.9  & 84.8 & 99.7 & 48.6 & \textbf{23.3} & 72.5 & 16.5 & 61.7 & 79.3 & 50.9 & 42.1 & 59.4 \\
PSELDNets & 26.1 & 54.5 & 59.3 & \textbf{57.5} & 24.2 & 35.0 & 15.9 & 63.1 & 101.0 & \textbf{75.2} & \textbf{23.0} & \textbf{23.9} \\
\textbf{Spatial-Dasheng}    & \textbf{31.6} & \textbf{42.5} & \textbf{44.3} & 54.1 & 24.9 & \textbf{34.6} & \textbf{26.4} & \textbf{53.8} & \textbf{78.4} & 72.2 & 24.3 & 24.7 \\
\bottomrule
\end{tabular}
\end{adjustbox}
\end{table}

%% file: table/02_spatial_benchmarks.tex
\begin{table}[t]
\caption{Results on the spatial-audio subset of MMAU-Pro and the SR subset of STAR-Bench. In addition to each benchmark's official metrics, we report channel-swap ACR to measure robustness to channel-order perturbations.}
% \caption{Results on the spatial-audio subset of MMAU-Pro and the SR subset of STAR-Bench. In addition to the official metrics, e.g., the full-set score and spatial accuracy for MMAU-Pro, and spatial accuracy and option-swap ACR for STAR-Bench, we report channel-swap ACR on both benchmarks to evaluate robustness to channel-order perturbations.}
\centering
\label{tab: spatial_benchmarks}
\begin{adjustbox}{width=\columnwidth,center}
\begin{tabular}{lcccccccc}
\toprule
\multirow{3}{*}{Model} & \multirow{3}{*}{Size} & \multirow{3}{*}{\makecell[c]{Audio\\Format}} & \multicolumn{3}{c}{MMAU-Pro} & \multicolumn{3}{c}{STAR-Bench (SR)}\\
\cmidrule(lr){4-6}\cmidrule(lr){7-9}
& & & \makecell[c]{Full-set\\Score} & \makecell[c]{Spatial\\Acc.} & \makecell[c]{Chan.-Swap\\ACR} & \makecell[c]{Spatial\\Acc.} & \makecell[c]{Opt.-Swap\\ACR} & \makecell[c]{Chan.-Swap\\ACR}\\
\midrule
Random Guess  & - & - & 23.4 & 21.2 & 13.5 & 33.3 & 3.7 & 11.1 \\
Human & - & - & 77.9 & 88.2 & - & 73.7 & - & - \\
\midrule
Gemini-3.6-Flash & - & Mono & \textbf{70.4} & 48.3 & 12.8 & 47.0 & 23.1 & 10.5 \\
MiMo-V2.5        & - & Mono & 65.1 & 43.1 & 14.2 & 43.8 & 12.4 & 16.9 \\
\addlinespace[0.5pt]
\hdashline
\addlinespace[1.5pt]
Qwen3-Omni-30B-A3B-Instruct & 30B & Mono & 60.6 & 36.6 & 0.0 & 44.7 & 17.3 & 0.0 \\
Qwen2.5-Omni-7B & 7B & Mono & 52.2 & 41.2 & 0.0 & 37.3 & 12.0 & 0.0 \\
MOSS-Audio-8B-Instruct & 8B & Mono & 57.5 & 26.8 & 0.0 & 41.7 & 8.2 & 0.0 \\
Audio-Flamingo-Next-Instruct & 8B & Mono & 56.9 & 37.2 & 0.7 & 24.2 & 7.4 & 0.0 \\
BAT & 7B & Binaural & 24.8 & 23.7 & 7.4 & 0.0 & 0.0 & 0.0 \\
MiDashengLM-7B-1021 & 8B & Mono & 55.9 & 18.2 & 0.0 & 44.3 & 20.3 & 0.0 \\
\textbf{MiDashengLM-Spatial-7B} & 8B & Binaural & 56.4 & \textbf{55.4} & \textbf{33.1} & \textbf{70.3} & \textbf{64.1} & \textbf{72.6} \\
\bottomrule
\end{tabular}
\end{adjustbox}
\vspace{-5mm}
\end{table}

%% file: table/03_general_benchmarks.tex
\begin{table}
\caption{General audio understanding results across audio question answering, ASR and audio captioning. In the last column, each cell reports our method and two ablation variants: (a) our full model, (b) ours with bidirectional connections in the audio encoder, and (c) ours without monaural corpora, arranged in the order a $|$ b $|$ c. The best and second-best results in each row are in bold and underlined, respectively.}
\centering
\label{tab: general_benchmarks}
\footnotesize
\begin{adjustbox}{width=\columnwidth,center}
\begin{tabular}{llccccc}
\toprule
Benchmark & Metric & \makecell{Qwen2.5-\\Omni-7B} & \makecell{MOSS-Audio\\-8B-Instruct} & \makecell{Audio-Flamingo\\-Next-Instruct} & \makecell{MiDashengLM\\-7B-1021} & \makecell{\textbf{MiDashengLM}\\\textbf{-Spatial-7B}} \\
\midrule
\multicolumn{7}{c}{\textit{Audio Question Answering}} \\
\addlinespace[0.5pt]
\hdashline
\addlinespace[1.5pt]
MMAU-Pro          & ACC   & 52.2 & \textbf{57.5} & \underline{56.9} & 55.9 & 56.4 $|$ 55.4 $|$ 50.9 \\
MMAU-v05.15.25    & ACC   & 71.5 & \textbf{76.7} & 71.4 & \underline{74.9} & 74.8 $|$ 73.4 $|$ 71.2 \\
MuChoMusic        & ACC   & 64.8 & 70.0 & \underline{75.6} & 73.0 & \textbf{76.7} $|$ 75.4 $|$ 70.1 \\
MusicQA           & FENSE & \underline{60.6} & 59.3 & 57.2 & \textbf{61.6} & 56.8 $|$ 56.9 $|$ 53.7 \\
AudioCaps-QA      & FENSE & \underline{53.3} & 45.6 & 49.8 & \textbf{54.2} & 52.5 $|$ 51.9 $|$ 51.5 \\
\midrule
\multicolumn{7}{c}{\textit{Automatic Speech Recognition}} \\
\addlinespace[0.5pt]
\hdashline
\addlinespace[1.5pt]
LibriSpeech (English) & & & & & & \\
\quad - \textit{test-clean}   & WER & \underline{1.7}  & 2.3 & \textbf{1.5} & 3.6  & 1.8 $|$ 1.9 $|$ 2.1 \\
\quad - \textit{test-other}   & WER & \underline{3.4}  & 5.6 & \textbf{2.8} & 5.9  & 4.2 $|$ 4.3 $|$ 5.2 \\
AISHELL-2 (Chinese) & & & & & & \\
\quad - \textit{Mic}          & CER & \textbf{2.5} & \underline{2.9} & 8.7 & 3.2  & 3.1 $|$ 3.2 $|$ 3.9 \\
\quad - \textit{iOS}          & CER & \textbf{2.6} & 3.0 & 8.3 & \underline{2.9} & \underline{2.9} $|$ 3.0 $|$ 3.6 \\
\quad - \textit{Android}      & CER & \textbf{2.7}  & \underline{3.0} & 7.5 & 3.1  & 3.2 $|$ 3.2 $|$ 4.2 \\
GigaSpeech2 & & & & & & \\
\quad - \textit{Indonesian}   & WER & 21.2 & 51.0 & 36.4 & 22.3 & \underline{20.9} $|$ \textbf{20.1} $|$ 34.3 \\
\quad - \textit{Thai}         & WER & 53.8 & 66.0 & $>100$ & 38.4 & \underline{37.4} $|$ \textbf{36.9} $|$ 62.8 \\
\quad - \textit{Vietnamese}   & WER & 18.6 & 40.1 & 59.9 & 17.7 & \textbf{16.2} $|$ \underline{16.9} $|$ 78.9 \\
\midrule
\multicolumn{7}{c}{\textit{Audio Captioning}} \\
\addlinespace[0.5pt]
\hdashline
\addlinespace[1.5pt]
MusicCaps      & FENSE & 43.7 & 50.6 & 56.2 & 59.1 & \textbf{60.0} $|$ \underline{59.9} $|$ 53.8 \\
Songdescriber  & FENSE & 45.3 & \underline{54.4} & \textbf{55.4} & 46.4 & 48.4 $|$ 47.9 $|$ 50.3 \\
AudioCaps      & FENSE & 60.8 & 46.4 & \textbf{64.0} & 62.1 & \underline{62.8} $|$ 61.8 $|$ 55.9 \\
ClothoV2       & FENSE & 47.6 & 37.8 & 48.6 & \underline{49.4} & \textbf{49.7} $|$ 48.4 $|$ 44.2 \\
AutoACD        & FENSE & 55.9 & 43.8 & 53.2 & \textbf{67.1} & 66.0 $|$ \underline{66.2} $|$ 58.2 \\
\bottomrule
\end{tabular}
\end{adjustbox}
\end{table}

%% file: tex/06_conclusion.tex
\section{Conclusion}
We present MiDashengLM-Spatial, the first open-source end-to-end unified audio-language model that supports both general audio understanding and spatial awareness within a single architecture. It extends MiDashengLM with Spatial-Dasheng, a spatial audio encoder pre-trained with a SELD objective for spatial perception, and integrates the two audio encoders through a hierarchical semantic-to-spatial conditioning (HSSC) module with minimal architectural modifications. We further develop a data synthesis pipeline that renders diverse spatial acoustic scenes along with scene-level descriptions and question-answer pairs to support training for spatial understanding. Experimental results show that Spatial-Dasheng exhibits strong spatial perception capability and generalizes well from simulated to real-world acoustic scenes. Building upon Spatial-Dasheng, MiDashengLM-Spatial achieves superior performance across spatial audio understanding and reasoning benchmarks, without compromising general audio understanding: it remains on par with state-of-the-art 8B-scale LALMs on monaural audio question answering, ASR, and audio captioning. These results show that general audio understanding and spatial awareness can coexist within a single model, marking a step toward human-like comprehensive auditory perception in audio-language models.

%% file: appendix/01_dataset_details.tex
\section{Dataset Details}
\subsection{Mapping from Spatial Attributes to Textual Descriptions}
\label{appendix: mapping}

\begin{figure}[h]
    \centering
    \includegraphics[width=\linewidth]{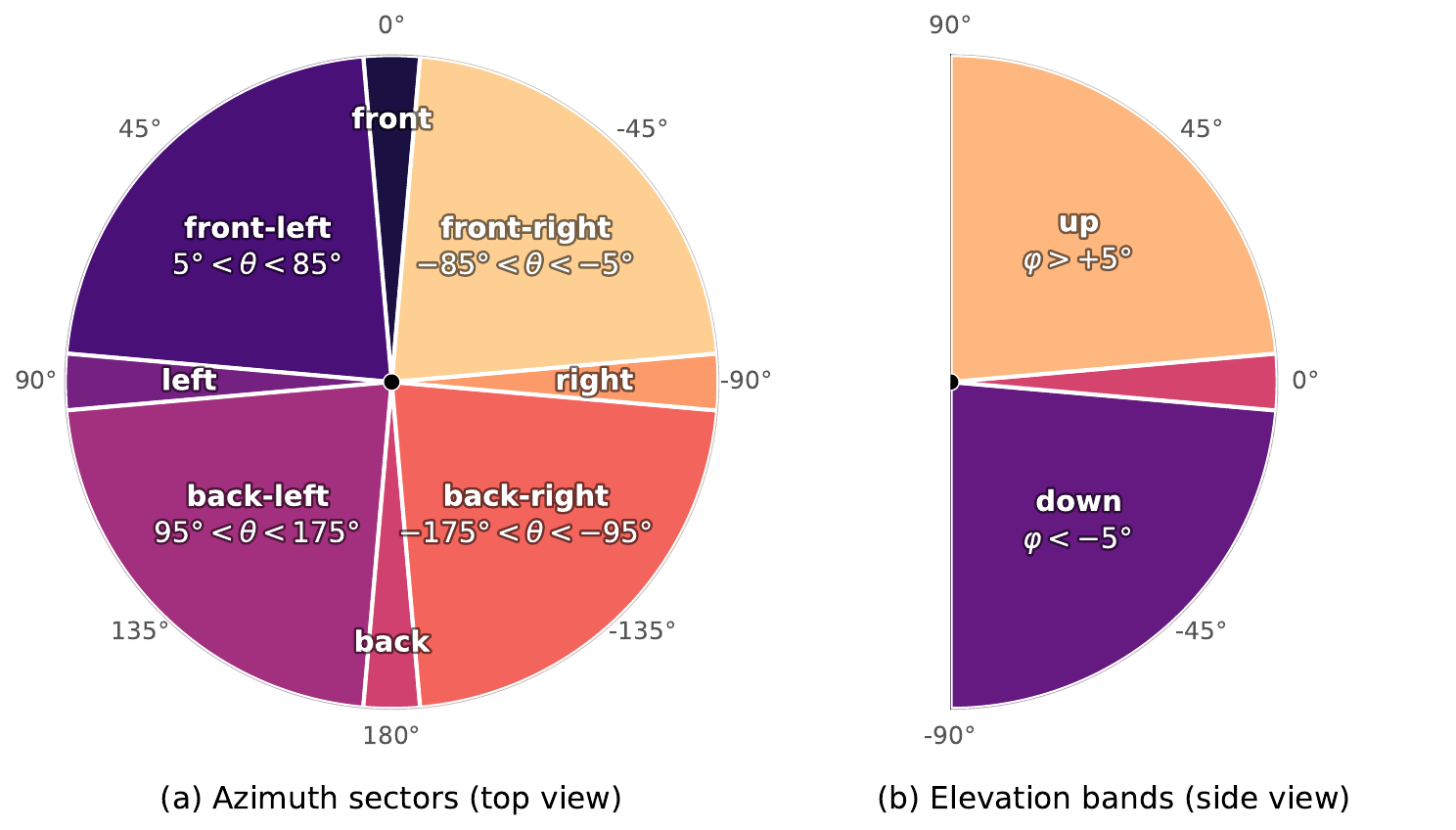}
    \caption{Mapping from spatial attributes to textual descriptions. (a) azimuth sectors (top view) and (b) elevation bands (side view), both defined in listener-centered coordinates. Sources with $|\phi| \leq 5^\circ$ are treated as horizontal and described without the \emph{up} or \emph{down} term.}
    \label{fig: direction_mapping}
\end{figure}

As part of the data pipeline (Section~\ref{sec: spatial_caption_gen}), we convert the numerical spatial attributes of each source into natural-language descriptors, as illustrated in Figure~\ref{fig: direction_mapping}. Specifically, the Cartesian source position $(x, y, z)$ is first transformed into spherical coordinates, i.e., azimuth $\theta \in [-180^\circ, +180^\circ]$ and elevation $\phi \in [-90^\circ, +90^\circ]$, in a listener-centered reference frame where positive azimuth points to the listener's left and positive elevation points upward. Subsequently, the azimuth is quantized into textual direction labels according to the sectors in Figure~\ref{fig: direction_mapping}~(a). The four cardinal direction labels \emph{front}, \emph{back}, \emph{left}, and \emph{right} are reserved for sources whose azimuth lies within $\pm 5^\circ$ of the corresponding axis (e.g., $|\theta| \leq 5^\circ$ for \emph{front}), while all remaining azimuths are assigned to the four diagonal sectors, each spanning $80^\circ$ (e.g., \emph{front-left} for $5^\circ < \theta < 85^\circ$). Sources with small elevation ($|\phi| \leq 5^\circ$) are treated as horizontal and described without the \emph{up} or \emph{down} term; otherwise, the \emph{up} or \emph{down} term is appended according to the elevation bands in Figure~\ref{fig: direction_mapping}~(b). For a moving scene, its trajectory is described by the motion direction together with the quantized start and end positions (e.g., \emph{moves counter-clockwise from front-right to front-left}).

\subsection{Scene-level Spatial Description}
\label{appendix: spatial_caption}

We generate scene-level spatial descriptions for both sound and speech scenes under a unified principle: all spatial and temporal facts are computed deterministically from the ground-truth metadata, and the LLM performs only the linguistic realization of these verified facts.

For sound scenes, the LLM builds a single coherent scene-level description of what happens, where it happens, and how events are temporally related. For speech scenes, we instead produce a time-ordered paragraph transcribing who says what, in which each speaker is assigned a deterministic referent derived from the metadata (e.g., \emph{the microphone wearer} for the wearer's own voice or \emph{the person on the listener's left} for a far-field speaker). The verbatim transcripts are never exposed to the LLM; they are substituted for the \texttt{<CONTENT\_i>} placeholders after a rephrasing pass that is accepted only if every placeholder is preserved intact and in order. The prompts used for generating scene-level spatial descriptions, along with example outputs, are presented below.

\input{box/01_prompt_spatial_audio_caption}

\input{box/02_prompt_spatial_speech_transcription}

\input{box/03_examples_spatial_audio_caption}

\input{box/04_examples_spatial_speech_transcription}

\subsection{Spatial Question-Answer Pairs}
\label{appendix: spatial_qa}

Section~\ref{sec: spatial_caption_gen} describes the overall construction of our spatial QA pairs. Table~\ref{tab: spatial_qa_templates} lists the question templates for the three deterministically constructed categories; each item is multiple-choice with three or four options. For spatially grounded content understanding, questions whose answer is a single speaker (i.e., \emph{who said X}) are converted into multiple-choice items whose options are the directions of all participants in the scene, while the remaining questions stay open-ended. Finally, an LLM rephrases each QA pair for linguistic diversity, including the question, options, and answer.

\input{table/04_spatial_qa_templates}

\input{box/05_qa_examples_source_localization}

\input{box/06_qa_examples_motion_trajectory}

\input{box/07_qa_examples_group_relations}

\input{box/08_qa_examples_content_understanding}

% \subsection{Dataset Statistics}
\label{appendix: data_stat}

%% file: box/01_prompt_spatial_audio_caption.tex
\begin{tcolorbox}[enhanced jigsaw, breakable,
  colback=blue!2!white,colframe=black!75!white,title=Prompt used for generating scene-level spatial audio captions,
  drop fuzzy shadow,watermark color=white,watermark text=,
  fontupper=\fontsize{7}{7.3}\selectfont,top=1pt,bottom=1pt,left=3pt,right=3pt]

You are an expert annotator for spatial audio scenes containing static and moving sources.
You do not hear the audio. You are given a chronological list of audio events that occur within a single audio recording.
For each event, you are provided with:\\
- an audio description\\
- start and end times (in seconds)\\
- a spatial information label (relative to the listener)\\
- an `Overlaps with' field listing which other events this event temporally overlaps with (or `none' if it does not overlap with any other event)

The spatial information label uses the listener as the reference point. For each event it may describe:\\
- a static source at a single direction, or\\
- a moving source with a starting direction, an ending direction, and a movement direction (e.g. ``clockwise'', ``counter-clockwise'', ``left-to-right'', ``right-to-left'', ``top-to-bottom'', ``bottom-to-top''), and\\
- an optional distance-change phrase, ``approaching the listener'' or ``moving away from the listener'', appearing only when the source's distance to the listener changes meaningfully, and\\
- optional stationary phases before or after the movement.

Important: the movement direction describes the trajectory itself; do NOT assume the trajectory is centered on the listener (do not say ``around the listener'' or ``circling the listener'' unless that is literally what the label states).

\smallskip
\textbf{Task:}
Produce a cohesive, fluent English paragraph describing the complete spatial audio scene. You must synthesize the individual events into a single narrative that accurately reflects both the spatial behaviour (static positions, motion trajectories, distance changes, stationary phases) and the temporal relationships (sequential, overlapping, or simultaneous) of the sounds.

\smallskip
\textbf{Instructions:}

1. Output Format \& Length Constraint: Output EXACTLY ONE final paragraph of 2-6 sentences, keeping it under 200 words. Do NOT include any headings, bullet points, or filler text.

2. Spatial Information is Mandatory: Each input event MUST be explicitly mentioned in the output, and each mentioned event MUST preserve its spatial information. For static events, state the direction. For moving events, preserve every component of the motion (initial direction, motion direction wording, final direction, distance-change phrase if present, and any stationary phase). Use the listener as the only spatial reference point.

3. Preserve Movement Wording: Keep the literal movement-direction wording from the label (e.g., ``clockwise'', ``counter-clockwise'', ``left-to-right'', ``top-to-bottom''). Do not rename or paraphrase the trajectory direction.

4. Semantic Compression: Reduce each event's audio description to a brief gist (typically 3-12 words). Include ONLY:\\
- the sound source type (e.g., man, woman, dog, instrument),\\
- the high-level action or category (e.g., speech, singing, sound effect),\\
- the language if non-English (e.g., German, Spanish),\\
- at most ONE highly distinctive cue per event if crucial.\\
Drop all detailed adjectives, audio-quality remarks, and redundant context.

5. Handling Overlaps: Use each event's `Overlaps with' field as the AUTHORITATIVE signal for temporal overlap --- do not infer overlap from start/end times alone. When the field lists one or more events, merge them into one sentence using terms like ``at the same time'', ``concurrently'', ``alongside'', or ``overlapping with''; you MUST STILL state the full spatial behaviour for each distinct event within that merged sentence. When the field is `none', the event does not overlap with anything else --- describe it sequentially (e.g. ``followed by'', ``after which'') and do NOT use overlap/simultaneity wording for it.

6. Cohesion and Fluency: Use natural conjunctions and transitional phrases to connect events. Do not produce a flat list.

7. Temporal Accuracy: Use start and end times to infer chronological ordering only. The `Overlaps with' field is the source of truth for whether two events temporally intersect; do not rely on time arithmetic for overlap decisions. Describe events as following one another when their `Overlaps with' fields do not list each other. Use expressions like ``starts with'', ``followed by'', ``overlapping with'', ``while'', ``simultaneously'', or ``throughout the recording''.

8. No Hallucination: Do not invent distance, motion, room layout, acoustic properties, or content beyond what the inputs state. In particular:\\
- Do not assert that a source orbits or circles the listener.\\
- Do not mention any change in distance, closeness, or loudness unless a distance-change phrase is supplied.

9. No Raw Timestamps or Coordinates: Do NOT use timestamp numbers or coordinates. Translate all timing into natural language.

10. Direction semantics:\\
- front = in front of the listener\\
- back = behind the listener\\
- left = to the listener's left\\
- right = to the listener's right\\
- up = above the listener\\
- down = below the listener\\
- directly = straight in the stated direction without additional offset unless explicitly provided\\
- clockwise / counter-clockwise = rotational direction of the trajectory as viewed from above (the trajectory is not necessarily centered on the listener)\\
- left-to-right / right-to-left / top-to-bottom / bottom-to-top / front-to-back / back-to-front = linear movement direction\\
- approaching the listener = the source's distance to the listener decreases noticeably during the movement\\
- moving away from the listener = the source's distance to the listener increases noticeably during the movement

\smallskip
\textbf{Example Input:}

Event 1:\\
- Description: A woman in English describes a shower's tiling and waterfall feature in a low-quality recording.\\
- Times: [1.8, 11.8]\\
- Spatial information: static at the listener's back, right, down\\
- Overlaps with: Event 2

Event 2:\\
- Description: A man speaks English about a child being accused of wrongdoing, expressing legal uncertainty.\\
- Times: [4.4, 14.4]\\
- Spatial information: moving left-to-right from the listener's front, left to front, right, up\\
- Overlaps with: Event 1

Event 3:\\
- Description: A German narrator describes biological specimens in a reverberant room.\\
- Times: [15.0, 24.8]\\
- Spatial information: initially stationary at the listener's front, right, up; moving clockwise from front, right, up to back, right, approaching the listener\\
- Overlaps with: none

\smallskip
\textbf{Example Output:}

The scene starts with a woman speaking from the listener's back-right and below, soon overlapping with a man talking that moves left-to-right from the listener's front-left to front-right and above. After both fade, a German narrator takes over, initially in front of the listener and to the right and above, then moving clockwise to the back-right while approaching the listener.

\end{tcolorbox}

%% file: box/02_prompt_spatial_speech_transcription.tex
\begin{tcolorbox}[enhanced jigsaw, breakable,
  colback=blue!2!white,colframe=black!75!white,title=Prompt used for generating scene-level spatial speech transcriptions,
  drop fuzzy shadow,watermark color=white,watermark text=,
  fontupper=\fontsize{7}{7.3}\selectfont,top=1pt,bottom=1pt,left=3pt,right=3pt]

You are an editor improving the wording of spatial speech-transcription descriptions for scenes containing static and moving speakers.

You are given a draft paragraph that narrates, in time order, who speaks in a multi-speaker scene and what each person says. Each spoken utterance has already been replaced by a placeholder of the form \texttt{<CONTENT\_0>}, \texttt{<CONTENT\_1>}, \texttt{<CONTENT\_2>}, ... in the order it occurs. Your job is to rewrite ONLY the framing sentence(s) around these placeholders so the paragraph reads naturally and varied, while keeping the exact same meaning.

Each speaker is referred to in one of three ways:\\
- A static far-field speaker is referred to by their fixed direction relative to the listener (e.g. ``the person on the front-left'', ``the speaker behind the listener'', ``the person directly to the right'').\\
- A moving far-field speaker is referred to by their full trajectory relative to the listener (e.g. ``the person moving clockwise from the listener's front-left to back-left'', ``the person moving left-to-right from the listener's front-left to front-right and approaching the listener'', ``the person initially stationary at the listener's back-right, then moving top-to-bottom to back-right-down'').\\
- The microphone wearer's own voice (the listener speaking) is referred to as ``the microphone wearer'', ``the listener'', or ``the person wearing the microphone''.

\smallskip
\textbf{Rules:}

- Keep EVERY placeholder token EXACTLY as written (\texttt{<CONTENT\_0>}, \texttt{<CONTENT\_1>}, ...). Do NOT add, remove, reorder, renumber, merge, split, or alter any placeholder. The number of placeholders in your output MUST equal the number in the input, each appearing exactly once.

- Preserve the temporal order of the placeholders. \texttt{<CONTENT\_0>} must still come before \texttt{<CONTENT\_1>}, and so on.

- SPATIAL INFORMATION must stay PRECISE. You may only change phrasing, never the spatial meaning itself.\\
\quad - For a STATIC speaker, keep the exact axis and side. Allowed rewrites:\\
\quad\quad ``the person on the front-left'' $\to$ ``the speaker to your front-left'' / ``the voice coming from your front-left''\\
\quad\quad ``the person directly in front of the listener'' $\to$ ``the speaker straight ahead of you''\\
\quad Never widen, narrow, flip, or drop an axis (do NOT turn ``front-left'' into ``left'', ``front'', or ``front-right'').\\
\quad - For a MOVING speaker, you MUST preserve EVERY component of the trajectory: the starting direction, the literal movement-direction wording, the ending direction, any distance-change phrase, and any stationary phase. Allowed rewrites change only phrasing, e.g.:\\
\quad\quad ``the person moving clockwise from the listener's front-left to back-left'' $\to$ ``the speaker travelling clockwise from your front-left round to your back-left''\\
\quad\quad ``the person moving left-to-right from the listener's front-left to front-right and approaching the listener'' $\to$ ``the voice that crosses left-to-right from your front-left to your front-right while drawing nearer''\\
\quad Keep the literal movement-direction wording (``clockwise'', ``counter-clockwise'', ``left-to-right'', ``right-to-left'', ``top-to-bottom'', ``bottom-to-top'', ``front-to-back'', ``back-to-front''). Do NOT rename or paraphrase the trajectory direction, drop the start or end direction, flip the order of start and end, or omit/invent a distance-change or stationary phase.

- Do NOT assume a moving trajectory is centered on the listener: do NOT say ``around the listener'', ``circling the listener'', or ``orbiting you'' unless the draft literally says so.

- Do NOT add a distance change (approaching / moving away / closer / louder) unless the draft already states one. Do NOT add or remove a stationary phase.

- Keep the microphone-wearer / listener self-voice clearly distinct from the far-field speakers. Do NOT assign the wearer a direction or a trajectory, and do NOT turn a directional or moving speaker into the wearer.

- Vary the connective wording so the sentences do not all start the same way, but keep the paragraph concise and natural. PRESERVE the temporal relationship each connective conveys:\\
\quad - A SEQUENTIAL connective (``then'', ``after that'', ``next'') means the utterance starts after the previous one finished. Keep it sequential (you may swap among ``then'' / ``after that'' / ``next'' / ``later'').\\
\quad - A CONCURRENT connective (``meanwhile'', ``at the same time'', ``simultaneously'', ``while that continues'', ``overlapping with that'') means the utterance is spoken WHILE an earlier one is still ongoing. Keep it concurrent (you may swap among these). Do NOT turn a concurrent connective into a sequential one, or vice versa --- that would misrepresent whether the speakers overlap.

- Do NOT describe, paraphrase, summarize, quote, or guess the spoken content. The placeholders stand in for the exact words and must remain untouched.

- Do NOT add new information: no loudness, room, timing, onset/offset, or audio quality unless it already appears in the draft.

- Do NOT use positional placeholders like ``Event 0'', ``Speaker 1'', or ``the first event'' to refer to a speaker; refer to each speaker only by direction/trajectory or as the microphone wearer.

\smallskip
\textbf{Direction semantics:}\\
- front = in front of the listener; back = behind the listener; left = to the listener's left; right = to the listener's right; up = above the listener; down = below the listener.\\
- directly = straight in the stated direction without additional offset.\\
- clockwise / counter-clockwise = rotational direction of the trajectory as viewed from above (not necessarily centered on the listener).\\
- left-to-right / right-to-left / top-to-bottom / bottom-to-top / front-to-back / back-to-front = linear movement direction.\\
- approaching the listener = distance decreases during the movement; moving away from the listener = distance increases during the movement.

\smallskip
Output the rewritten paragraph as plain text only. Do not output markdown, comments, JSON, or any text other than the paragraph itself.

\smallskip
\textbf{Example Input:}

The person moving counter-clockwise from the listener's back-right-up to front-right-up says \texttt{<CONTENT\_0>}. Then the microphone wearer says \texttt{<CONTENT\_1>}. Next, the person on the left-down says \texttt{<CONTENT\_2>}.

\smallskip
\textbf{Example Output:}

The speaker travelling counter-clockwise from your back-right-up round to your front-right-up says \texttt{<CONTENT\_0>}. After that, the microphone wearer says \texttt{<CONTENT\_1>}, and then a voice from your lower left says \texttt{<CONTENT\_2>}.

\end{tcolorbox}

%% file: box/03_examples_spatial_audio_caption.tex
\begin{tcolorbox}[enhanced jigsaw, breakable,
  colback=blue!1!white,colframe=black!50!white,title=Examples of scene-level spatial audio captions,
  drop fuzzy shadow,watermark color=white,watermark text=,top=1pt,bottom=1pt]
  \setlength{\leftmargini}{0.1em}
  \begin{itemize}
    \item The recording begins with a woman giving a speech, static at the listener's front, right, and above. This is followed by a machine working, static at the listener's back, right, and below, and then a cat meowing, static at the listener's front, right, and above. Finally, a door slams from a static position at the listener's front, left, and above.
    \item The recording begins with a speaker detailing shroud installation moving counter-clockwise from the listener's back, right to front, right, overlapping with a man discussing school start times who is static at the listener's front, right, up. This is followed by a speaker discussing quartz, static at the listener's front, left, up, occurring concurrently with a man explaining light bar wiring, static at the listener's front, right, up. This man's explanation also overlaps with a reporter narrating a shooting moving left-to-right from the listener's back, down to back, right, down. Finally, the reporter's narrative overlaps with a woman describing eyeshadow application, static at the listener's back, left, up.
    \item The scene begins with an engine starting, moving counter-clockwise from the listener's front-left-down to back-right-down and moving away from the listener, concurrently with a vehicle motor revving static at the listener's front-left-down and a sewing machine working static at the listener's back-right-down. This is followed by a motor revving moving counter-clockwise from the listener's back-left-up to front-right-up, overlapping with water running moving left-to-right from the listener's front-left-down to front-right-down and a train horn moving clockwise from the listener's back-right-up to front-left-up. The train horn then overlaps with a helicopter flying, moving clockwise from the listener's front-right-down to back-left-down, alongside an accelerating sound static at the listener's front-right-down and rapid tapping moving back-to-front from the listener's back-right to directly right and approaching the listener.
  \end{itemize}
\end{tcolorbox}

%% file: box/04_examples_spatial_speech_transcription.tex
\begin{tcolorbox}[enhanced jigsaw, breakable,
  colback=blue!1!white,colframe=black!50!white,title=Examples of scene-level spatial speech transcriptions,
  drop fuzzy shadow,watermark color=white,watermark text=,top=1pt,bottom=1pt]
  \setlength{\leftmargini}{0.1em}
  \begin{itemize}
    \item The speaker moving clockwise within the listener's front, right, down area says ``I never saw any one yet that did not look simply and out of countenance nor ever knew a wedding well designed but one and that was of two persons who had time enough I confess to contrive it and nobody to please in't but themselves''. Then, a voice from above and on your back-right says ``Major Foster in a letter to Judge George M Bennett of Minneapolis said''. Following that, the microphone wearer says ``And I declare it's too bad that it is'', and then the person wearing the microphone says ``And he placed it in that gentleman's fingers who now took his turn at the lamp and contemplated the little parallelogram with a gleam of sly amusement''.
    \item The voice coming from above and to your back-right says ``Well as I say it's an awful queer world they clap all the burglars into jail and the murderers and the wife beaters I've allers thought a gentle reproof would be enough punishment for a wife beater cause he probably has a lot o provocation that nobody knows and the firebugs can't think o the right name something like cendenaries an the breakers o the peace an what not an yet the law has nothin to say to a man like Hen Lord''. After that, the microphone wearer says ``The colonel did not disclose the fact that it was no news to him''.
    \item A voice from your back-right-up says ``Twenty people sings out''. Simultaneously, the speaker above and to your back-right says ``It is a chamber on the ground floor which has a grated window opening on the garden which is closed on the outside by a shutter and two doors one leads into the convent the other into the church what church''. Overlapping with that, the person below and to your front-left says ``In spite of myself I drew my breath in sharply'', while the microphone wearer says ``If for a whim you beggar yourself I cannot stay you''. While that continues, the voice from your front-left-down says ``Here I found a great ship ready for sea and full of merchants and notables who had with them goods of price so I embarked my bales therein''. Meanwhile, the person moving top-to-bottom within your back, right, up area and approaching the listener says ``Oh she's always at the piano said Van she must be there now somewhere and then somebody laughed''. Next, the speaker above and to your back-right says ``It was specially suited for a trial plant also in the early days when a yield of six or eight lamps to the horse power was considered subject for congratulation''. Overlapping with that, the person moving back-to-front within your front, left, down area says ``This outward mutability indicated and did not more than fairly express the various properties of her inner life''. After that, the microphone wearer says ``It was in fact the best weapon of its day'', followed by the person above and on your back-right saying ``But I have occasionally done the same thing at other times''.
  \end{itemize}
\end{tcolorbox}

%% file: table/04_spatial_qa_templates.tex
\begin{table}[h]
\caption{Categories and question-answer templates of the deterministically constructed spatial QA pairs. Sound-scene and speech-scene QA share the same templates. \texttt{\{sound\}}: a short description of a sound source (e.g., \emph{a dog barking}); \texttt{\{speech\}}: a short summary of what a speaker says; \texttt{\{dir\}}: a quantized direction phrase (e.g., \emph{front-left}).}
\centering
\label{tab: spatial_qa_templates}
\footnotesize
\setlength{\tabcolsep}{4pt}
\begin{tabular}{@{}>{\centering\arraybackslash}p{2.6cm}p{10.7cm}@{}}
\toprule
Category & \multicolumn{1}{c}{Question-Answer Template} \\
\midrule
\multicolumn{2}{c}{\textit{Source Localization}} \\
\addlinespace[0.5pt]
\hdashline
\addlinespace[1.5pt]
\multirow{3}{*}{\makecell{Source-to-Direction\\ Localization}} &
   \textbf{Q}: From which direction is \texttt{\{sound\}} coming? / From which direction do you hear the person who says \texttt{\{speech\}}? \textbf{A}: \texttt{\{dir\}} \newline
   \textbf{Q}: Is \texttt{\{sound\}} above or below the listener? \textbf{A}: above / below / Level with the listener \\
\addlinespace[0.5pt]
\hdashline
\addlinespace[1.5pt]
\multirow{2}{*}{\makecell{Direction-to-Source\\ Identification}} &
   \textbf{Q}: Which sound comes from the \texttt{\{dir\}}? \textbf{A}: \texttt{\{sound\}}\newline
   \textbf{Q}: What does the person on your \texttt{\{dir\}} say? \textbf{A}: \texttt{\{speech\}}\\
\midrule
\multicolumn{2}{c}{\textit{Motion Trajectory Perception}} \\
\addlinespace[0.5pt]
\hdashline
\addlinespace[1.5pt]
\multirow{6}{*}{\makecell{Trajectory Tracking}} &
   \textbf{Q}: Does \texttt{\{sound\}} / the person who says \texttt{\{speech\}} move or stay in place?
   \textbf{A}: It moves / It stays in place / It cannot be determined\newline
   \textbf{Q}: How does \texttt{\{sound\}} / the person who says \texttt{\{speech\}} move? \textbf{A}: Clockwise / Counter-clockwise / Stays in place\newline
   \textbf{Q}: Where does \texttt{\{sound\}} / the person who says \texttt{\{speech\}} start out / end up? \textbf{A}: \texttt{\{dir\}} (the start/end position of the movement path) \\
\addlinespace[0.5pt]
\hdashline
\addlinespace[1.5pt]
\multirow{2}{*}{\makecell{Distance Change\\ Tracking}} &
   \textbf{Q}: Does \texttt{\{sound\}} / the person who says \texttt{\{speech\}} move closer to or farther from the listener (you)? \textbf{A}: Getting closer / Moving away / Distance unchanged \\
\midrule
\multicolumn{2}{c}{\textit{Group-level Spatio-temporal Relations}} \\
\addlinespace[0.5pt]
\hdashline
\addlinespace[1.5pt]
\multirow{6}{*}{\makecell{Multi-Source\\Spatial Relation}} &
   \textbf{Q}: Are \texttt{\{sound\}} and \texttt{\{sound\}} on the same side (left or right) of the listener? \textbf{A}: Yes / No / Uncertain \newline
   \textbf{Q}: Which sound is more to the left / further in front / closer to the listener? 
   \textbf{A}: \texttt{\{sound\}} \newline
   \textbf{Q}: On the left-right (or front-back) axis, where is \texttt{\{sound\}} relative to \texttt{\{sound\}}? \textbf{A}: To the left / To the right / To the front / To the back / In the same place \\
\addlinespace[0.5pt]
\hdashline
\addlinespace[1.5pt]
\multirow{2}{*}{\makecell{Spatial Counting}} &
   \textbf{Q}: How many sound sources are on the \texttt{\{dir\}} side / How many people speak from your \texttt{\{dir\}} throughout the scene? \textbf{A}: 0 / 1 / 2 /... \\
\addlinespace[0.5pt]
\hdashline
\addlinespace[1.5pt]
\multirow{2}{*}{\makecell{Temporal Order}} &
   \textbf{Q}: Which person speaks first / last? \textbf{A}: The person who says \texttt{\{speech\}}\newline
   \textbf{Q}: From which direction does the first / last utterance in the scene come? \textbf{A}: \texttt{\{dir\}} \\
\bottomrule
\end{tabular}
\end{table}

%% file: box/05_qa_examples_source_localization.tex
\begin{tcolorbox}[enhanced jigsaw, breakable,
  colback=blue!1!white,colframe=black!50!white,title=Spatial QA examples for Source Localization,
  drop fuzzy shadow,watermark color=white,watermark text=,top=1pt,bottom=1pt]
  \setlength{\leftmargini}{0.1em}
  \begin{itemize}
    \item \textbf{Q}: In which direction is a distorted vehicle engine running unevenly located? \{To your front-right; Directly to your left; To your back-right; To your front-left\} \newline
    \textbf{A}: To your front-left
    \item \textbf{Q}: Which sound is located to your front-left? \{The cheerful electronic track; The slow meditative acoustic guitar piece; The French-speaking woman; The Tamil-speaking man\} \newline
    \textbf{A}: The Tamil-speaking man
  \end{itemize}
\end{tcolorbox}

%% file: box/06_qa_examples_motion_trajectory.tex
\begin{tcolorbox}[enhanced jigsaw, breakable,
  colback=blue!1!white,colframe=black!50!white,title=Spatial QA examples for Motion Trajectory Perception,
  drop fuzzy shadow,watermark color=white,watermark text=,top=1pt,bottom=1pt]
  \setlength{\leftmargini}{0.1em}
  \begin{itemize}
    \item \textbf{Q}: Does an easygoing piano-driven jazz piece stay in one place or does it move? \{It moves; It cannot be determined; It stays in place\} \newline
    \textbf{A}: It moves
    \item \textbf{Q}: In what way does an excited man discussing the significance of a sports trophy move? \{Moving from right to left; Moving from left to right; Staying in place\} \newline
    \textbf{A}: Moving from right to left
    \item \textbf{Q}: Does the slow melancholic instrumental piano piece move closer to you or further away? \{Moving away; Distance unchanged; Getting closer\} \newline
    \textbf{A}: Getting closer
  \end{itemize}
\end{tcolorbox}

%% file: box/07_qa_examples_group_relations.tex
\begin{tcolorbox}[enhanced jigsaw, breakable,
  colback=blue!1!white,colframe=black!50!white,title=Spatial QA examples for Group-level Spatio-temporal Relations,
  drop fuzzy shadow,watermark color=white,watermark text=,top=1pt,bottom=1pt]
  \setlength{\leftmargini}{0.1em}
  \begin{itemize}
    \item \textbf{Q}: Which of these sounds is located further in front? \{The Japanese-speaking woman; The meditative piano piece; The Spanish-speaking woman\} \newline
    \textbf{A}: The meditative piano piece
    \item \textbf{Q}: How many speakers are positioned behind you throughout the scene? \{3; 2; 1; 0\} \newline
    \textbf{A}: 1
    \item \textbf{Q}: From which direction do you hear the very first utterance in the scene? \{To your back-right; To your front-left; The microphone wearer; To your back-left\} \newline
    \textbf{A}: To your front-left
  \end{itemize}
\end{tcolorbox}

%% file: box/08_qa_examples_content_understanding.tex
\begin{tcolorbox}[enhanced jigsaw, breakable,
  colback=blue!1!white,colframe=black!50!white,title=Spatial QA examples for Spatially Grounded Content Understanding,
  drop fuzzy shadow,watermark color=white,watermark text=,top=1pt,bottom=1pt]
  \setlength{\leftmargini}{0.1em}
  \begin{itemize}
    \item \textbf{Q}: Who mentions Taylor Swift? \{The person above and to your front-left; The person below and to your front-right\} \newline
    \textbf{A}: The person below and to your front-right
    \item \textbf{Q}: Where does the person above and to your front-right currently live? \newline
    \textbf{A}: China
    \item \textbf{Q}: What is the favorite Mexican dish of the person directly behind you? \newline
    \textbf{A}: Enchiladas
  \end{itemize}
\end{tcolorbox}

%% file: appendix/02_experiment_details.tex
\section{Details for Spatially Grounded Evaluation}

\subsection{Spatial Audio-Text Alignment Evaluation}
\label{appendix: spatial_align_eval}

\subsubsection{Evaluation Metrics}
\label{appendix: spatial_grounded_metrics}

To evaluate whether the spatial information expressed in generated text is grounded in the input audio, we employ an LLM-assisted structured evaluation approach. Conventional text-generation metrics, such as FENSE \cite{fense}, CIDEr \cite{vedantam2015cider}, and SPICE \cite{spice}, assess semantic similarity but do not explicitly measure spatial or temporal correctness, whereas standard SELD metrics \cite{overviewofDCASE} require structured predictions and cannot be applied directly to open-ended text. We use Gemma-4-31B to parse each generated scene- or event-level description into a temporally ordered set of sound events, represented in JSON format by their semantic content, direction, motion state, and potential motion trajectory. These structured predictions are then matched with the reference events and evaluated both separately and jointly in terms of their semantic, spatial, and temporal consistency.

\begin{figure}[p]
  \centering
  \begin{minipage}[t]{0.5\linewidth}
    \centering
    \includegraphics[width=0.9\linewidth]{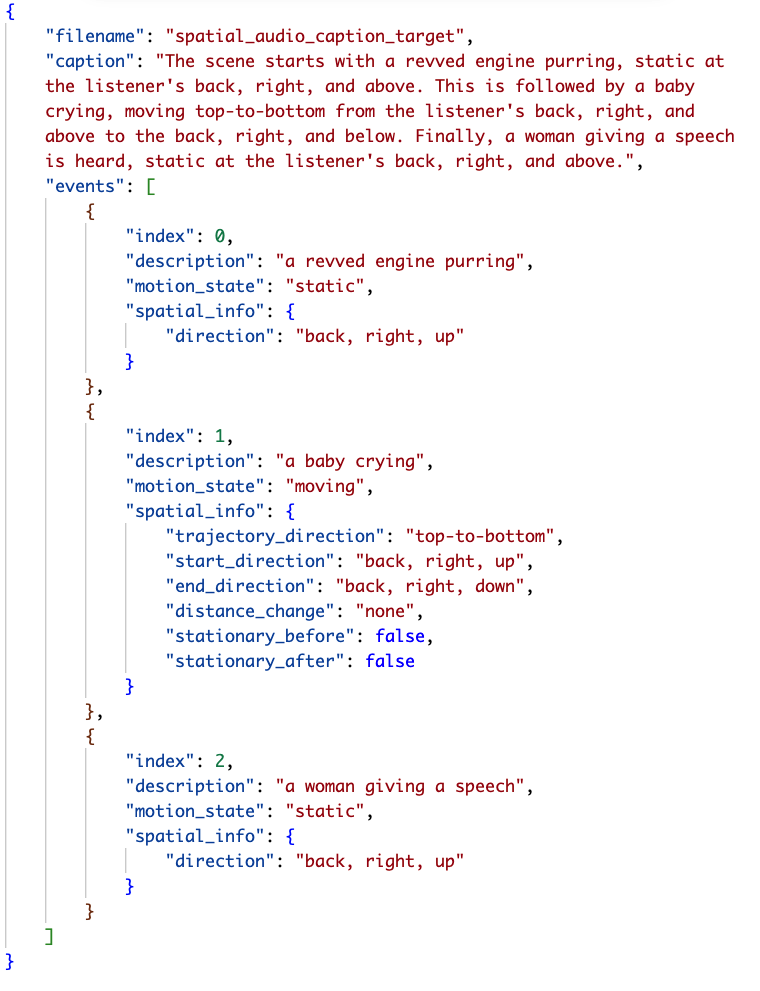}
    \centerline{\small (a) Reference spatial audio caption}
  \end{minipage}\hfill
  \begin{minipage}[t]{0.5\linewidth}
    \centering
    \includegraphics[width=0.9\linewidth]{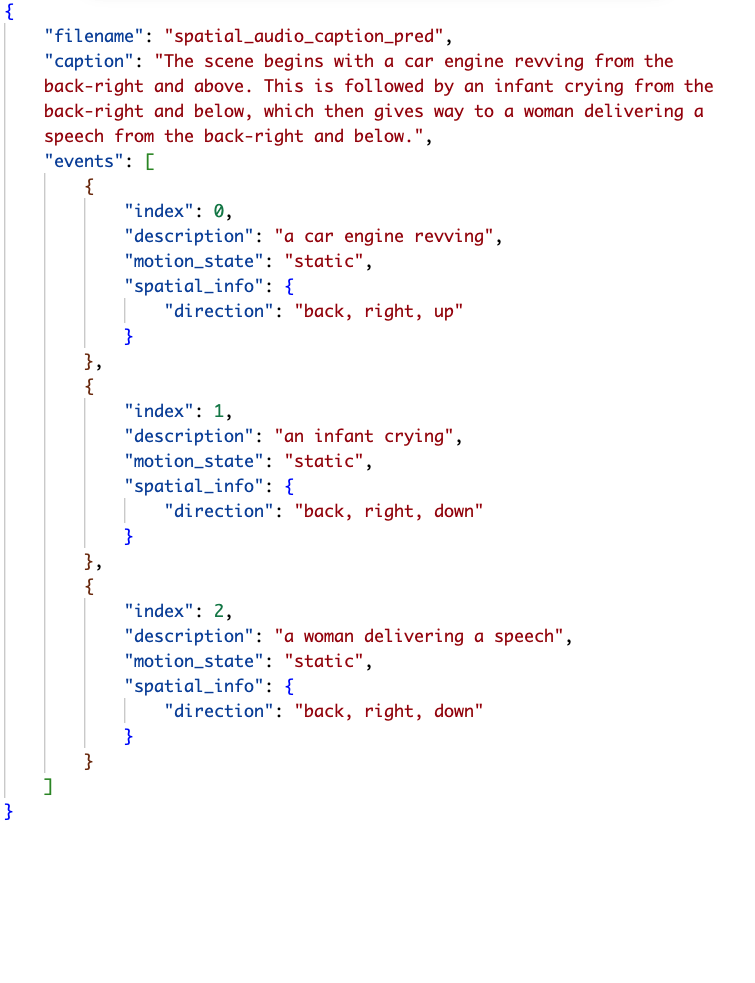}
    \centerline{\small (b) Predicted spatial audio caption}
  \end{minipage}
  \begin{minipage}[t]{0.5\linewidth}
    \centering
    \includegraphics[width=0.9\linewidth]{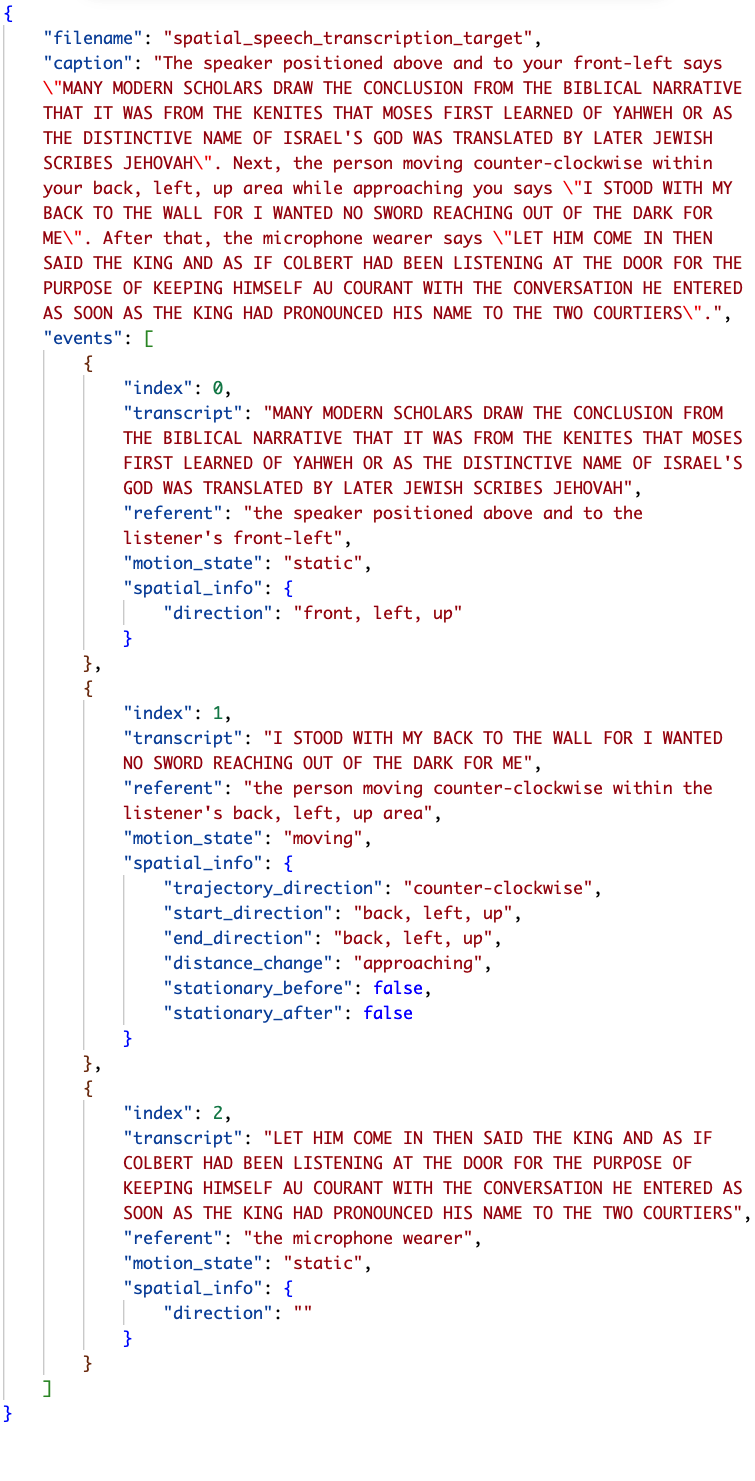}
    \centerline{\small (c) Reference spatial speech transcription}
  \end{minipage}\hfill
  \begin{minipage}[t]{0.5\linewidth}
    \centering
    \includegraphics[width=0.9\linewidth]{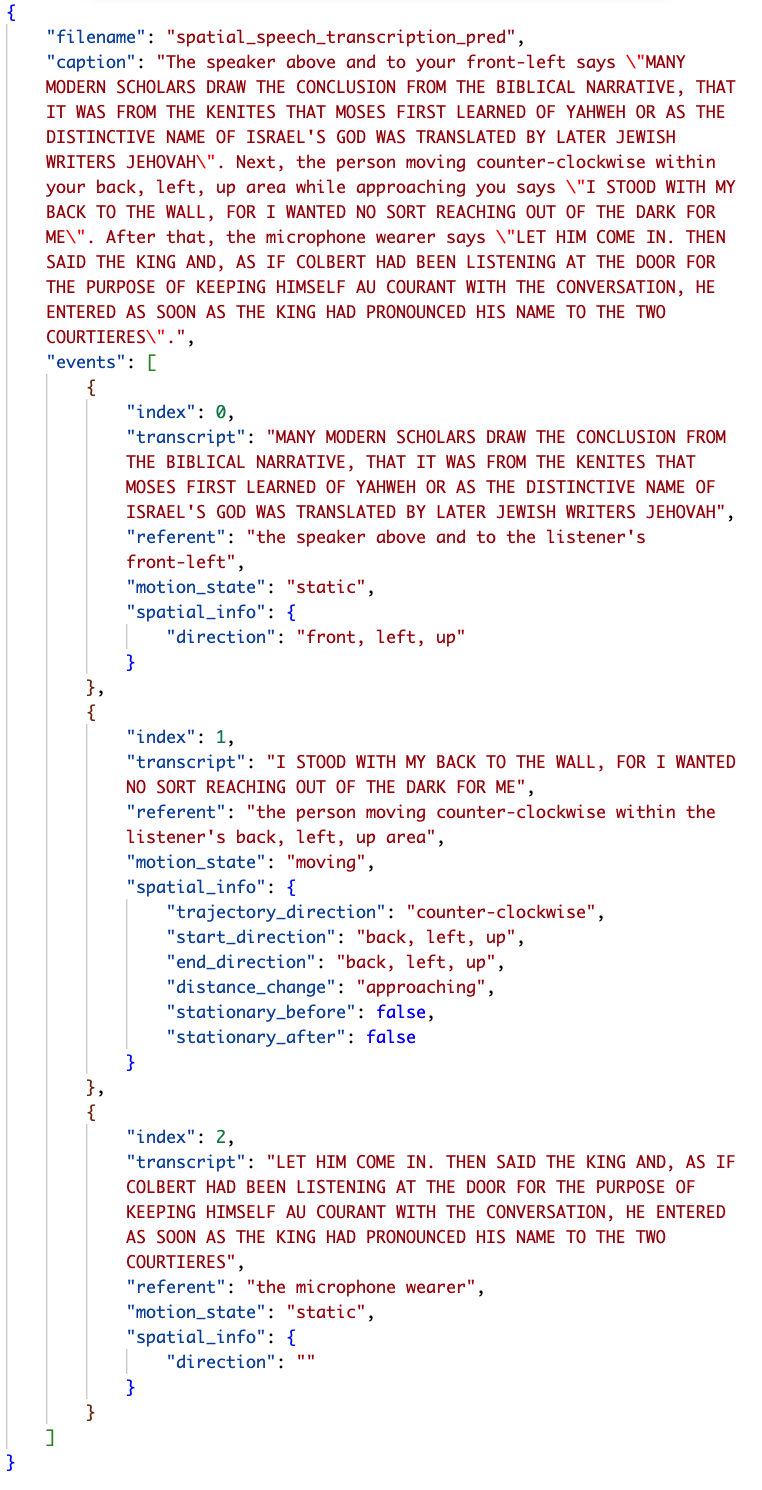}
    \centerline{\small (d) Predicted spatial speech transcription}
  \end{minipage}
  \caption{Examples of reference and predicted outputs for spatial audio captioning and spatial speech transcription, together with the structured attributes reverse-parsed into JSON format by an LLM.}
  \label{fig: spatially_grounded_examples}
\end{figure}

Figure \ref{fig: spatially_grounded_examples} presents examples of spatial audio captions and spatial speech transcriptions, together with their parsed per-event representations, including semantic descriptions (or transcriptions), the corresponding temporal order, and spatial attributes. These examples demonstrate how open-ended textual outputs are converted into a common structured representation on which the metrics below are computed.

Spatial audio captioning and spatial speech transcription share this evaluation procedure and differ only in the measure used to establish semantic correspondence between predicted and reference events: FENSE~\cite{fense} is used for captions, whereas word error rate (WER) is used for transcripts. Event correspondence is established independently within each scene. Once this correspondence has been established, the same spatial-attribute and temporal-order scoring procedures are applied to both tasks and aggregated over the evaluation set. Algorithm~\ref{alg:scene_metrics} summarizes the complete evaluation approach.

\input{table/05_scene_metrics_algorithm}

\paragraph{Metric definitions.}
Let $\mathcal{S}$ denote the set of evaluation scenes. For each scene $s\in\mathcal{S}$, semantic matching between its reference events $R_s$ and predictions $H_s$ yields the scene-level match set $\mathcal{M}_s\subseteq R_s\times H_s$. Let $N_{\mathrm{match}}=\sum_{s\in\mathcal{S}}|\mathcal{M}_s|$ and $N_{\mathrm{ref}}=\sum_{s\in\mathcal{S}}|R_s|$ denote the total numbers of matched pairs and reference events, respectively, with $N_{\mathrm{ref},s}=|R_s|$. $N_{\mathrm{match}}^{\mathrm{stat}}$ and $N_{\mathrm{match}}^{\mathrm{mov}}$ denote the total numbers of matched static and moving events whose motion states are correctly predicted across all evaluation scenes, respectively.
\setlength{\leftmargini}{1em}
\begin{itemize}
  \item \textbf{Recall}: $N_{\mathrm{match}}/N_{\mathrm{ref}}$, the fraction of
        reference events with a semantic match.
  \item \textbf{Motion State}: $\sum_{s\in\mathcal{S}}\sum_{(i,j)\in\mathcal{M}_s}m_{s,ij}/N_{\mathrm{match}}$, the motion-state (static or moving) accuracy over all matched events.
  \item \textbf{Static Match}: $\sum_{s\in\mathcal{S}}\sum_{(i,j)\in\mathcal{M}_s}s^{\mathrm{stat}}_{s,ij}/N_{\mathrm{match}}^{\mathrm{stat}}$, the direction accuracy over all matched static events whose motion state is correctly predicted.
  \item \textbf{Moving Match}: $\sum_{s\in\mathcal{S}}\sum_{(i,j)\in\mathcal{M}_s}s^{\mathrm{mov}}_{s,ij}/N_{\mathrm{match}}^{\mathrm{mov}}$, the full-trajectory (e.g., trajectory, start/end position, distance change, stationary before/after) accuracy over all matched moving events whose motion state is correctly predicted.
  \item \textbf{Temporal Order}: $\sum_{s\in\mathcal{S}} \sum_{\{(i,j),(i',j')\}\subseteq\mathcal{M}_s}p_{s,ii'}\big/\sum_{s\in\mathcal{S}} \frac{N_{\mathrm{ref},s}(N_{\mathrm{ref},s}-1)}{2}$, a Kendall-style pairwise concordance score defined as the fraction of correctly ordered reference-event pairs. A pair is correct if both events are semantically matched and either their predicted narration order is consistent with the reference onset order or the corresponding onset difference is within $\delta$.
  \item \textbf{Overall}: $\sum_{s\in\mathcal{S}}\sum_{(i,j)\in\mathcal{M}_s}s_{s,ij}z_{s,ij}/N_{\mathrm{ref}}$, the fraction of reference events that are semantically matched, spatially correct ($s_{s,ij}=1$), and temporally concordant ($z_{s,ij}=1$), where $s_{s,ij}$ denotes the applicable static or moving spatial indicator and is zero for an incorrect motion state.
\end{itemize}

\subsubsection{Results}

\input{table/06_spatial_grounded}

Table \ref{tab: spatial_grounded} presents the spatially grounded evaluation results on synthetic spatial audio-text datasets, comparing predictions from the MiDashengLM-Spatial model trained through Stage II with oracle scores derived from ground-truth spatial descriptions. The oracle results indicate that the JSON-based parser reliably extracts semantic content, motion states, source directions, and temporal order, while showing slightly lower accuracy for motion trajectories, which involve more diverse and compositional attributes. The relatively low \emph{Overall} scores across subsets reflect the strict requirement that all evaluated attributes are correct simultaneously. In scenes with non-overlapping sources, MiDashengLM-Spatial identifies most sound events ($\mathrm{Recall}>70\%$), accurately infers their motion states ($\mathrm{Motion~State}>88\%$) and static positions ($\mathrm{Static~Match}>50\%$), and reliably determines the temporal relationships between events ($\mathrm{Temporal~Order}>96\%$). Performance generally declines as scene complexity increases, particularly in the presence of overlapping or moving sources. Overall, these results suggest that MiDashengLM-Spatial achieves promising spatially grounded cross-modal alignment and is more effective at capturing coarse-grained spatial cues, such as motion states, than fine-grained attributes, e.g., motion trajectories.

\subsection{Analysis of Differences in Localization Accuracy}
\label{appendix: localization_prf_diff}

\input{table/07_localization_prf}

To determine whether integrating Spatial-Dasheng into MiDashengLM preserves its localization capability, we compare Spatial-Dasheng with MiDashengLM-Spatial after Stage II on the synthetic test subset containing non-overlapping static sources. Because Spatial-Dasheng predicts frame-level coordinates whereas MiDashengLM-Spatial generates event-level textual directions, we convert both outputs into a common text-based representation. Specifically, the Spatial-Dasheng coordinates are averaged over the active frames of each event and mapped to textual directions according to Figure \ref{fig: direction_mapping}. Table \ref{tab: localization_prf} reports precision, recall, and F-score along the front-back, left-right, and up-down axes, together with an \textit{Overall} exact-match score that requires all three axes to be correct.

Table \ref{tab: localization_prf} shows that the integration does not degrade the localization information learned by Spatial-Dasheng. Instead, MiDashengLM-Spatial improves the three-axis exact-match F-score from $44.7\%$ to $50.8\%$. This improvement suggests that end-to-end event-level prediction can consolidate spatial evidence over the duration of an event and may hence be less sensitive to inconsistencies among relatively independently estimated frame-level positions. Moreover, both models perform better along the left-right axis than along the other two axes. The result is consistent with human binaural perception.

\subsection{Spatial Audio Question-Answering Evaluation}
\label{appendix: spatial_aqa_eval}

\input{table/08_spatial_qa_results}

Table \ref{tab: spatial_qa_results} reports category-level accuracy on the synthetic spatial QA test sets defined in Table \ref{tab: spatial_qa_templates}. Each model score is accompanied by the corresponding random-guess accuracy. The evaluated MiDashengLM-Spatial model was trained through Stage III using only the multichannel corpus, without any additional monaural training data, to isolate its performance under multichannel-only training. OV1 contains scenes without source overlap, whereas OV2 contains two overlapping sources and, as a consequence of the data construction, more sound events and a larger amount of scene-level context.

MiDashengLM-Spatial substantially outperforms random guessing in every category, achieving overall accuracies of $80.6\%$ on OV1 and $66.9\%$ on OV2. The decline on OV2 indicates that spatial reasoning becomes more difficult when acoustic cues from concurrent sources are entangled, and more events must be represented jointly. Spatial Counting is the most challenging category, with accuracy decreasing from $65.2\%$ to $39.5\%$, the largest drop among the seven categories. Unlike tasks that query a single source or a local relation, Spatial Counting requires the model to identify and spatially distinguish relevant events throughout the scene before aggregating them into a single answer. Source overlap makes event-level identification and separation more difficult, while the larger number of events increases both the counting burden and the amount of context that must be integrated across the scene.

%% file: table/05_scene_metrics_algorithm.tex
\begin{figure}[h]
\begin{minipage}{\linewidth}
\begin{algorithm}[H]
\DontPrintSemicolon
\caption{Unified evaluation procedure for spatial captioning and transcription.}
\label{alg:scene_metrics}
\KwIn{Evaluation scenes $\mathcal{S}$, with reference events $R_s=\{r_{s,i}\}$ and predictions $H_s=\{h_{s,j}\}$ for each $s\in\mathcal{S}$; match threshold $\theta$ (FENSE $\geq 0.5$ for spatial audio captioning; WER $\leq 20\%$ for spatial speech transcription); onset tolerance $\delta=1$\,s}
\KwOut{Recall, Motion State, Static Match, Moving Match, Temporal Order, Overall}

\BlankLine
\tcp{Step 1: Content-based one-to-one matching within each scene}
\ForEach{$s\in\mathcal{S}$}{
  \ForEach{$(i,j)\in R_s\times H_s$}{
    \uIf(\tcp*[f]{spatial audio caption}){caption}{
      $C^{(s)}_{ij} \leftarrow \mathrm{FENSE}\big(\text{caption}\ r_{s,i},\ \text{caption}\ h_{s,j}\big)$
      \tcp*{maximize}
    }
    \Else(\tcp*[f]{spatial speech transcription}){
      $C^{(s)}_{ij} \leftarrow \mathrm{WER}\big(\text{transcript}\ r_{s,i},\ \text{transcript}\ h_{s,j}\big)$
      \tcp*{minimize}
    }
  }
  $\widetilde{\mathcal{M}}_s \leftarrow \mathrm{Hungarian}(C^{(s)})$\;
  $\mathcal{M}_s \leftarrow \{(i,j)\in\widetilde{\mathcal{M}}_s:C^{(s)}_{ij}\ \mathrm{passes}\ \theta\}$
  \tcp*{semantically correct matches}
}

\BlankLine
\tcp{Step 2: Spatial scoring over within-scene matched pairs}
\ForEach{$s\in\mathcal{S}$}{
  \ForEach{$(i,j)\in\mathcal{M}_s$}{
    $m_{s,ij} \leftarrow \mathds{1}[\mathrm{motion}\ r_{s,i} = \mathrm{motion}\ h_{s,j}]$
    \tcp*{Motion State}

    \If{$m_{s,ij}=1$}{
      \uIf{$\mathrm{motion}\ r_{s,i}\ \mathbf{is}$ static}{
        $s^{\mathrm{stat}}_{s,ij} \leftarrow \mathds{1}[\mathrm{direction}\ r_{s,i} = \mathrm{direction}\ h_{s,j}]$
        \tcp*{Static Match}
      }
      \Else{
        $s^{\mathrm{mov}}_{s,ij} \leftarrow \mathds{1}[\mathrm{trajectory}\ r_{s,i} = \mathrm{trajectory}\ h_{s,j}]$
        \tcp*{Moving Match}
      }
    }
  }
}

\BlankLine
\tcp{Step 3: Temporal order over within-scene matched pairs}
\ForEach{$s\in\mathcal{S}$}{
  $t_{s,i} \leftarrow \mathrm{onset}(r_{s,i})$;\ 
  $q_{s,j} \leftarrow \mathrm{index}(h_{s,j})$
  \tcp*{start time and narration order}
  $z_{s,ij} \leftarrow 1,\ \forall (i,j)\in\mathcal{M}_s$
  \tcp*{event-level temporal concordance}
  \ForEach{$(i,j),(i',j')\in\mathcal{M}_s\ \mathrm{such\ that}\ i<i'$}{
    \uIf{$|t_{s,i}-t_{s,i'}|\leq\delta$}{
      $p_{s,ii'} \leftarrow 1$
      \tcp*{near-simultaneous events form a tie}
    }
    \Else{
      $p_{s,ii'} \leftarrow
      \mathds{1}\!\left[\mathrm{sgn}(t_{s,i}-t_{s,i'})
      =\mathrm{sgn}(q_{s,j}-q_{s,j'})\right]$
      \tcp*{order agrees}
      \If{$p_{s,ii'}=0$}{
        $z_{s,ij},\ z_{s,i'j'} \leftarrow 0$
        \tcp*{temporal order mismatch}
      }
    }
  }
}
\end{algorithm}
\noindent$\mathds{1}[\cdot]$ is the indicator function.
\end{minipage}
\end{figure}

%% file: table/06_spatial_grounded.tex
\begin{table}[h]
\caption{The performance of spatial audio-text alignment on synthetic spatial audio captioning and spatial speech transcription datasets. In each cell, values are reported as \ora{Oracle} $\vert$ Model, where \ora{Oracle} denotes the upper-bound result computed directly from ground-truth spatial descriptions and Model denotes the corresponding predictions from MiDashengLM-Spatial.}
\centering
\label{tab: spatial_grounded}
\begin{adjustbox}{width=\columnwidth,center}
\begin{tabular}{lcccccc}
\toprule
Subset & Overall (\%) & Recall (\%) & Motion State (\%) & Moving Match (\%) & Static Match (\%) & Temporal Order (\%) \\
\midrule
\multicolumn{7}{c}{\textit{Spatial Audio Captioning}} \\
\addlinespace[0.5pt]
\hdashline
\addlinespace[1.5pt]
event / moving / ov1     & \ora{86.76} $\vert$ 30.86 & \ora{99.87} $\vert$ 71.25 & \ora{96.21} $\vert$ 89.57 & \ora{81.61} $\vert$ 15.67 & \ora{99.50} $\vert$ 52.21 & \ora{-} $\vert$ -             \\
scene / moving / ov1  & \ora{90.39} $\vert$ 21.40 & \ora{99.43} $\vert$ 73.20 & \ora{99.81} $\vert$ 90.95 & \ora{82.52} $\vert$ 12.84 & \ora{99.92} $\vert$ 50.69 & \ora{99.78} $\vert$ 96.14     \\
scene / moving / ov2  & \ora{89.90} $\vert$ 8.23  & \ora{99.38} $\vert$ 56.40 & \ora{99.63} $\vert$ 70.67 & \ora{82.62} $\vert$ 6.66  & \ora{99.62} $\vert$ 36.78 & \ora{99.64} $\vert$ 89.51     \\
% scene / moving / ov3  & \ora{89.19} $\vert$ 4.55  & \ora{99.36} $\vert$ 44.95 & \ora{99.44} $\vert$ 62.70 & \ora{83.16} $\vert$ 4.83  & \ora{99.56} $\vert$ 33.07 & \ora{99.51} $\vert$ 86.39     \\
\midrule
\multicolumn{7}{c}{\textit{Spatial Speech Transcription}} \\
\addlinespace[0.5pt]
\hdashline
\addlinespace[1.5pt]
event / moving / ov1      & \ora{97.90} $\vert$ 65.68 & \ora{100.00} $\vert$ 84.57 & \ora{100.00} $\vert$ 92.73 & \ora{91.56} $\vert$ 9.43 & \ora{100.00} $\vert$ 56.69 & \ora{-} $\vert$ -             \\
scene / moving / ov1  & \ora{97.84} $\vert$ 51.94 & \ora{99.89}  $\vert$ 74.11 & \ora{100.00} $\vert$ 88.45 & \ora{91.76} $\vert$ 9.08 & \ora{100.00} $\vert$ 55.91 & \ora{100.00} $\vert$ 100.00   \\
scene / moving / ov2  & \ora{94.98} $\vert$ 8.39  & \ora{98.13}  $\vert$ 9.63  & \ora{100.00} $\vert$ 79.26 & \ora{91.42} $\vert$ 2.13 & \ora{99.97}  $\vert$ 55.00 & \ora{100.00} $\vert$ 100.00   \\
\bottomrule
\end{tabular}
\end{adjustbox}
\end{table}

%% file: table/07_localization_prf.tex
\begin{table}[h]
\caption{Comparison of localization performance on the synthetic test subset containing non-overlapping static sources. Each cell reports precision, recall and F-score, in that order. \textit{Overall} denotes exact matches across all three spatial axes.}
\centering
\label{tab: localization_prf}

\begin{adjustbox}{width=0.9\columnwidth,center}
% \begin{tabular}{@{}ccccc@{}}
\begin{tabular}{ccccc}
\toprule
Model & Front-Back (\%) & Left-Right (\%) & Up-Down (\%) & Overall (\%) \\
\midrule
Spatial-Dasheng     & \textbf{82.4}$\,\vert\,$78.2$\,\vert\,$80.2          & 95.4$\,\vert\,$\textbf{94.0}$\,\vert\,$\textbf{94.7} & \textbf{78.5}$\,\vert\,$65.6$\,\vert\,$71.4          & 45.3$\,\vert\,$44.2$\,\vert\,$44.7                   \\
MiDashengLM-Spatial & 80.2$\,\vert\,$\textbf{82.1}$\,\vert\,$\textbf{81.1} & \textbf{95.6}$\,\vert\,$93.4$\,\vert\,$94.5          & 74.5$\,\vert\,$\textbf{77.3}$\,\vert\,$\textbf{75.9} & \textbf{51.2}$\,\vert\,$\textbf{50.4}$\,\vert\,$\textbf{50.8} \\
\bottomrule
\end{tabular}
\end{adjustbox}
\end{table}

%% file: table/08_spatial_qa_results.tex
\begin{table}[h]
\caption{Accuracy of the seven fine-grained categories on the synthetic spatial QA test sets. In each cell, values are reported as Model $\vert$ \ora{Random}, where Model and \ora{Random} denote the results of MiDashengLM-Spatial predictions and the random guess baseline, respectively.}
\centering
\label{tab: spatial_qa_results}
\begin{adjustbox}{width=0.6\columnwidth,center}
% \begin{tabular}{@{}lcc@{}}
\begin{tabular}{lcc}
\toprule
\multicolumn{1}{c}{Category} & OV1 (\%) & OV2 (\%) \\
\midrule
Source-to-Direction Localization  & 87.6 $\vert$ \ora{28.9} & 79.0 $\vert$ \ora{29.1} \\
Direction-to-Source Identification & 86.1 $\vert$ \ora{26.2} & 63.7 $\vert$ \ora{25.0} \\
Trajectory Tracking               & 82.7 $\vert$ \ora{30.0} & 68.2 $\vert$ \ora{29.8} \\
Distance Change Tracking          & 71.2 $\vert$ \ora{33.3} & 75.9 $\vert$ \ora{33.3} \\
Multi-Source Spatial Relation     & 83.5 $\vert$ \ora{33.3} & 71.5 $\vert$ \ora{33.3} \\
Spatial Counting                  & 65.2 $\vert$ \ora{25.0} & 39.5 $\vert$ \ora{25.0} \\
Temporal Order                    & 98.8 $\vert$ \ora{25.1} & 86.8 $\vert$ \ora{25.0} \\
\midrule
Overall                           & 80.6 $\vert$ \ora{28.8} & 66.9 $\vert$ \ora{28.8} \\
\bottomrule
\end{tabular}
\end{adjustbox}
\end{table}

%% file: appendix/03_contributors.tex
\section{Contributors}

\begin{minipage}[t]{0.30\textwidth}
  \centering
  \textbf{Core Contributors} \\
  Jinbo Hu \\
  Hang Su \\
  Lichun Fan \\
  
\end{minipage}
\hfill
\begin{minipage}[t]{0.30\textwidth}
  \centering
  \textbf{Contributors} \\
  Heinrich Dinkel \\
  Gang Li \\
  Zhanchen Dai \\
  Yiru Zhang \\
  Chang Liu \\
  Peng Wang \\
  Junnan Wu \\
\end{minipage}
\hfill
\begin{minipage}[t]{0.30\textwidth}
  \centering
  \textbf{Supervisors} \\
  Jian Luan \\
  Cong Zou \\
  Heng Qu \\
  
\end{minipage}